\documentclass[11pt]{article}

\usepackage{acl}

\usepackage{times}
\usepackage{latexsym}
\usepackage[T1]{fontenc}
\usepackage[utf8]{inputenc}
\usepackage{microtype}
\usepackage{inconsolata}
\usepackage{graphicx}
\graphicspath{{figures/}}
\usepackage{amsmath,amssymb}
\usepackage{booktabs}
\usepackage{multirow}
\usepackage{xcolor}
\usepackage{xspace}
\usepackage{pifont}
\usepackage{colortbl}
\usepackage{subcaption}
\usepackage{tikz}
\usepackage{pgfplots}
\pgfplotsset{compat=1.16}
\usepackage{algorithm}
\usepackage{algorithmic}
\usepackage{enumitem}
\usepackage[most]{tcolorbox}
\tcbuselibrary{listings,breakable}

\definecolor{systembg}{RGB}{232,240,254}   
\definecolor{systemborder}{RGB}{100,149,220}
\definecolor{userbg}{RGB}{255,249,230}     
\definecolor{userborder}{RGB}{210,160,50}
\definecolor{assistbg}{RGB}{232,248,238}   
\definecolor{assistborder}{RGB}{80,170,110}
\definecolor{prompttitlebg}{RGB}{50,80,140}

\tcbset{
  promptbase/.style={
    breakable, enhanced, fontupper=\small\ttfamily,
    left=4pt, right=4pt, top=3pt, bottom=3pt,
    boxrule=0.6pt, arc=3pt,
    before upper={\setlength{\parskip}{2pt}}
  },
  systemprompt/.style={
    promptbase,
    colback=systembg, colframe=systemborder,
    title={\sffamily\bfseries\footnotesize System},
    attach boxed title to top left={yshift=-2pt,xshift=4pt},
    boxed title style={colback=systemborder,colframe=systemborder,arc=2pt,
      left=2pt,right=2pt,top=1pt,bottom=1pt}
  },
  userprompt/.style={
    promptbase,
    colback=userbg, colframe=userborder,
    title={\sffamily\bfseries\footnotesize User},
    attach boxed title to top left={yshift=-2pt,xshift=4pt},
    boxed title style={colback=userborder,colframe=userborder,arc=2pt,
      left=2pt,right=2pt,top=1pt,bottom=1pt}
  },
  assistprompt/.style={
    promptbase,
    colback=assistbg, colframe=assistborder,
    title={\sffamily\bfseries\footnotesize Model Output (excerpt)},
    attach boxed title to top left={yshift=-2pt,xshift=4pt},
    boxed title style={colback=assistborder,colframe=assistborder,arc=2pt,
      left=2pt,right=2pt,top=1pt,bottom=1pt}
  },
  strategybox/.style={
    enhanced, breakable,
    colback=white, colframe=prompttitlebg,
    fonttitle=\sffamily\bfseries\small\color{white},
    attach boxed title to top left={yshift=-3pt,xshift=6pt},
    boxed title style={colback=prompttitlebg,arc=2pt,
      left=4pt,right=4pt,top=1pt,bottom=1pt},
    boxrule=1pt, arc=4pt,
    left=6pt, right=6pt, top=8pt, bottom=4pt
  }
}

\newcommand{\sys}{\textsc{AlgoBench}\xspace}

\newcommand{\optt}{\textsc{OptT}\xspace}
\newcommand{\opts}{\textsc{OptS}\xspace}
\newcommand{\traprate}{\textsc{TrapRate}\xspace}
\newcommand{\gapt}{\textsc{GapT}\xspace}
\newcommand{\consens}{\textsc{ConSens}\xspace}

\newcommand{\cmark}{\ding{51}}
\newcommand{\xmark}{\ding{55}}

\definecolor{csblue}{RGB}{30,100,180}
\definecolor{csorange}{RGB}{210,100,30}
\definecolor{csgreen}{RGB}{40,150,80}
\definecolor{lightgray}{RGB}{245,245,245}
\definecolor{lightblue}{RGB}{220,235,252}

\title{\textsc{AlgoBench}: Benchmarking Algorithmic Adaptation in Code Generation}

\author{\textbf{Xinyuan Song}$^{1}$ \quad
    \textbf{Zekun Cai}$^{2,3}$ \quad
    \textbf{Liang Zhao}$^{1}$ \\
    $^{1}$Emory University, Atlanta, GA, USA \quad
    $^{2}$The University of Tokyo, Tokyo, Japan \\
    $^{3}$LocationMind, Tokyo, Japan \\
    \texttt{\{xinyuan.song,liang.zhao\}@emory.edu, caizekun@csis.u-tokyo.ac.jp} \\
}

\begin{document}
\maketitle

\begin{abstract}
High pass rates on established programming benchmarks such as HumanEval and LiveCodeBench do not always show whether a model can reason about algorithms. Many fixed benchmarks eventually become part of the public training ecosystem through released problem statements, editorials, and generated solutions, allowing later models to improve partly by exposure rather than by stronger algorithmic ability. We introduce \sys, a framework that automatically builds \emph{novel} algorithmic problems from known competitive-programming problems through structured constraint-shifting transformations. Each accepted \sys variant is traceable to a source problem, but must make the original reference algorithm fail. Beyond pass@$k$, we introduce complexity-aware metrics---including \optt, \opts, \traprate, \gapt, and \consens---to test whether a solution is not only functionally correct but also asymptotically suitable for the generated problem. Experiments across multiple LLMs and prompting strategies show that performance drops sharply on \sys variants, retrieval can increase reuse of the old algorithm, and many correct-looking solutions fail to meet the required complexity. Error analysis shows that failures are mainly algorithmic rather than implementation-level, suggesting that \sys evaluates adaptation beyond functional correctness.
Code is available at \url{https://github.com/Hik289/algobench.git}.
\end{abstract}

\section{Introduction}
\label{sec:intro}
Large language models (LLMs) now perform well on many programming and algorithmic benchmarks. Recent systems obtain high pass@1 scores on standard datasets such as HumanEval \citep{chen2021evaluating} and LiveCodeBench \citep{jain2024livecodebench}. These scores, however, do not necessarily measure algorithmic reasoning. Many programming problems, editorials, and reference solutions are publicly available, and web-scale pretraining corpora may contain exact or near-duplicate problem--solution pairs. A model can therefore pass a benchmark by recalling a known solution pattern, rather than by deriving the required algorithm from the stated constraints \citep{golchin2024timetravelllms,shi2024detecting}.

Recent benchmarks reduce direct contamination by using newer or harder problems. LiveCodeBench \citep{jain2024livecodebench} and LiveBench \citep{white2024livebench} collect released tasks, while ProBench evaluates models on competitive-programming problems with online submissions, difficulty grading, and algorithm-tag analysis \citep{yang2025probench}. Humanity's Last Code Exam (HLCE) further uses IOI and ICPC World Finals problems to test advanced reasoning models on difficult contest tasks \citep{li2025hlce}. These benchmarks improve over older static datasets, but they remain fixed once released. Their problem statements, editorials, and model-generated solutions can later enter training corpora, allowing future models to improve partly by exposure rather than by stronger algorithmic reasoning. Thus, a benchmark for algorithmic ability should not rely only on a static set of hard or recently collected problems.

This motivates a different benchmark-design question: \textbf{how can we automatically generate new algorithmic problems so that LLMs cannot improve by memorizing existing problem statements and solutions?} Such a benchmark satisfy two conditions. First, each generated problem should be traceable to a known source problem, so that old-template reuse can be measured rather than only suspected. Second, the generated problem should require a new algorithmic treatment or a different asymptotic complexity, so that memorizing the source problem is not enough to solve.

\begin{figure*}[t]
  \centering
  \includegraphics[width=0.75\textwidth]{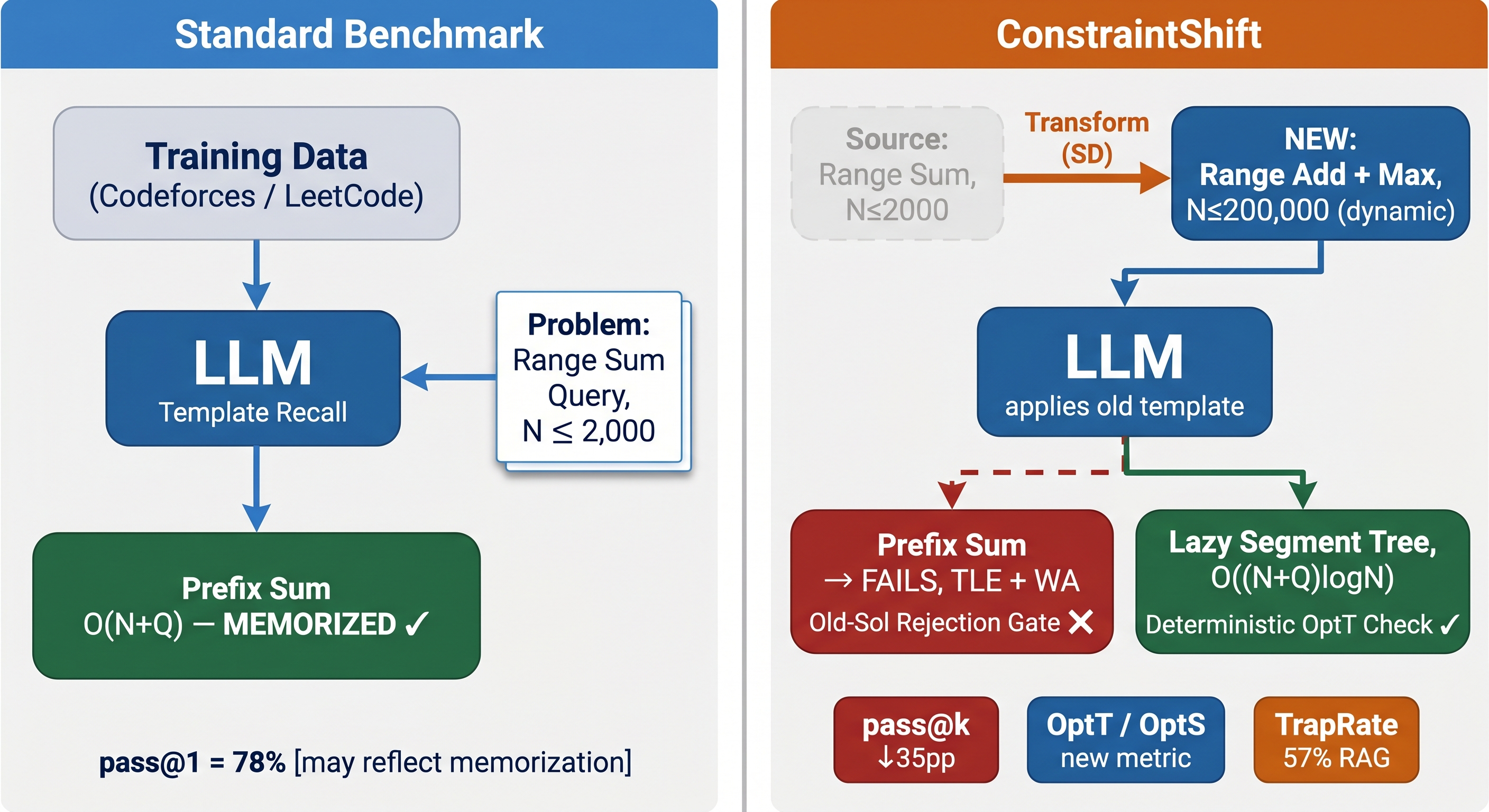}
  \caption{Core idea of \sys. A standard benchmark may allow an LLM to reuse a familiar template such as \emph{Prefix Sum}. \sys changes the problem so that the original solution fails under the new constraints, and then tests whether the model can produce the required new algorithm, such as \emph{Lazy Segment Tree}, with the correct asymptotic complexity.}
  \label{fig:intuition}
\end{figure*}

We propose \sys, a benchmark framework for automatically generating new algorithmic problems and evaluating algorithmic adaptation. As illustrated in Figure~\ref{fig:intuition}, \sys starts from competitive-programming problems with known reference algorithms. It then applies structured transformations, including constraint scaling, static-to-dynamic conversion, objective perturbation, and greedy-trap injection. These transformations produce new problems that remain traceable to their sources, but solving them requires algorithmic treatment rather than reuse of the original solution. Before a generated problem is added to the benchmark, it must pass four quality gates: the original reference solution must fail, a new reference solution must be verified, the statement must pass similarity filtering, and the target time and space complexity must be certified. This design also follows recent calls for benchmark quality control, reproducibility, and transparent validation in code-related LLM evaluation \citep{cao2025how2bench}.

The benchmark is designed to evaluate more than test passing. Standard pass@$k$ only checks whether a submitted program passes the tests; it does not separate an asymptotically suitable algorithm from a slower one that happens to pass under a permissive time limit. To make this distinction, \sys includes a deterministic three-layer complexity verifier. The verifier combines static AST-level analysis, algorithm-tag checks, and calibrated runtime scaling tests to assign auditable time and space optimality labels. This lets \sys measure whether a model produces a solution that is both correct and aligned with the required algorithmic complexity.

We evaluate seven LLMs under six prompting strategies on a 420-problem primary split drawn from 598 accepted \sys variants. The results show a clear drop from source problems to automatically generated variants, suggesting that many models struggle when memorized templates are no longer sufficient. Retrieval-augmented prompting, although useful in some settings, can also increase reuse of the source algorithm because the retrieved source problem anchors the model to the original solution. Error analysis further shows that most failures are algorithmic: old-solution reuse and too-slow algorithms are much more common than ordinary implementation bugs. These results suggest that \sys tests a different ability from standard code benchmarks: solving newly generated algorithmic problems that require adaptation beyond memorized templates.

Our contributions are summarized as follows:
\begin{itemize}[left = 0em,noitemsep]
  \item We introduce \sys, an automatic benchmark construction framework for generating new algorithmic problems and evaluating algorithms.
  \item We design ten rule-based transformation operators and quality gates to ensure each generated problem is valid, non-paraphrastic, traceable to a source problem, and rejects the original solution.
  \item We build a benchmark of 598 accepted algorithmic variants with metadata, reference solutions, and brute-force oracles; the main experiments use a 420-problem primary split.
  \item We propose complexity-aware metrics beyond pass@$k$, including \optt, \opts, \traprate, \gapt, and \consens.
  \item We evaluate seven LLMs under six prompting strategies and show that automatically generated problems reveal old-template reuse and suboptimal algorithmic reasoning.
\end{itemize}


\section{Problem Formulation}
\label{sec:formulation}

\paragraph{Source problem.}
We represent a source problem as
$q=(\textit{stmt}, \mathcal{C}, \alpha, T^*, S^*, a^*)$.
Here, $\textit{stmt}$ is the problem statement, including the input/output format and public examples. $\mathcal{C}$ is the constraint set, such as $n \leq 2000$. $\alpha$ is the reference algorithm type, such as \texttt{prefix\_sum}. $T^*$ and $S^*$ are the reference time and space complexities, and $a^*$ is the reference solution.
\paragraph{Generated problem.}
A generated problem is written as
$q' = T(q,\delta)$,
where $T$ is a transformation operator and $\delta$ specifies the concrete change. For example, $\delta$ may increase the input bound from $n\leq 2000$ to $n\leq 2\times10^5$, add update operations, change the objective, or introduce an additional constraint. The generated problem has its own target time complexity $\hat{T}^*$, target space complexity $\hat{S}^*$, algorithm type $\hat{\alpha}$, and reference solution $\hat{a}^*$.
\paragraph{Acceptance conditions.}
A generated problem $q'$ is accepted only if it satisfies all conditions below:
\begin{enumerate}[label=\textbf{C\arabic*.},leftmargin=*,noitemsep]
  \item $\text{TextSim}(q, q') < \tau_\text{text}$ (non-paraphrase)
  \item $q'$ has an unambiguous and deterministically judgeable specification (well-defined)
  \item Running $a^*$ on hidden tests for $q'$ produces WA, TLE, or MLE (old solution fails)
  \item A verified $\hat{a}^*$ exists and agrees with a brute-force oracle on inputs (new solution verified)
  \item $\hat{a}^*$ satisfies $\hat{T}^*$ and $\hat{S}^*$ according to the complexity verifier (complexity certified)
  \item $\hat{\alpha} \neq \alpha$ or $\hat{T}^* \neq T^*$ (algorithmic change)
\end{enumerate}

\section{Transformation Framework}
\label{sec:framework}
\begin{figure*}[t]
  \centering
  \includegraphics[width=0.75\textwidth]{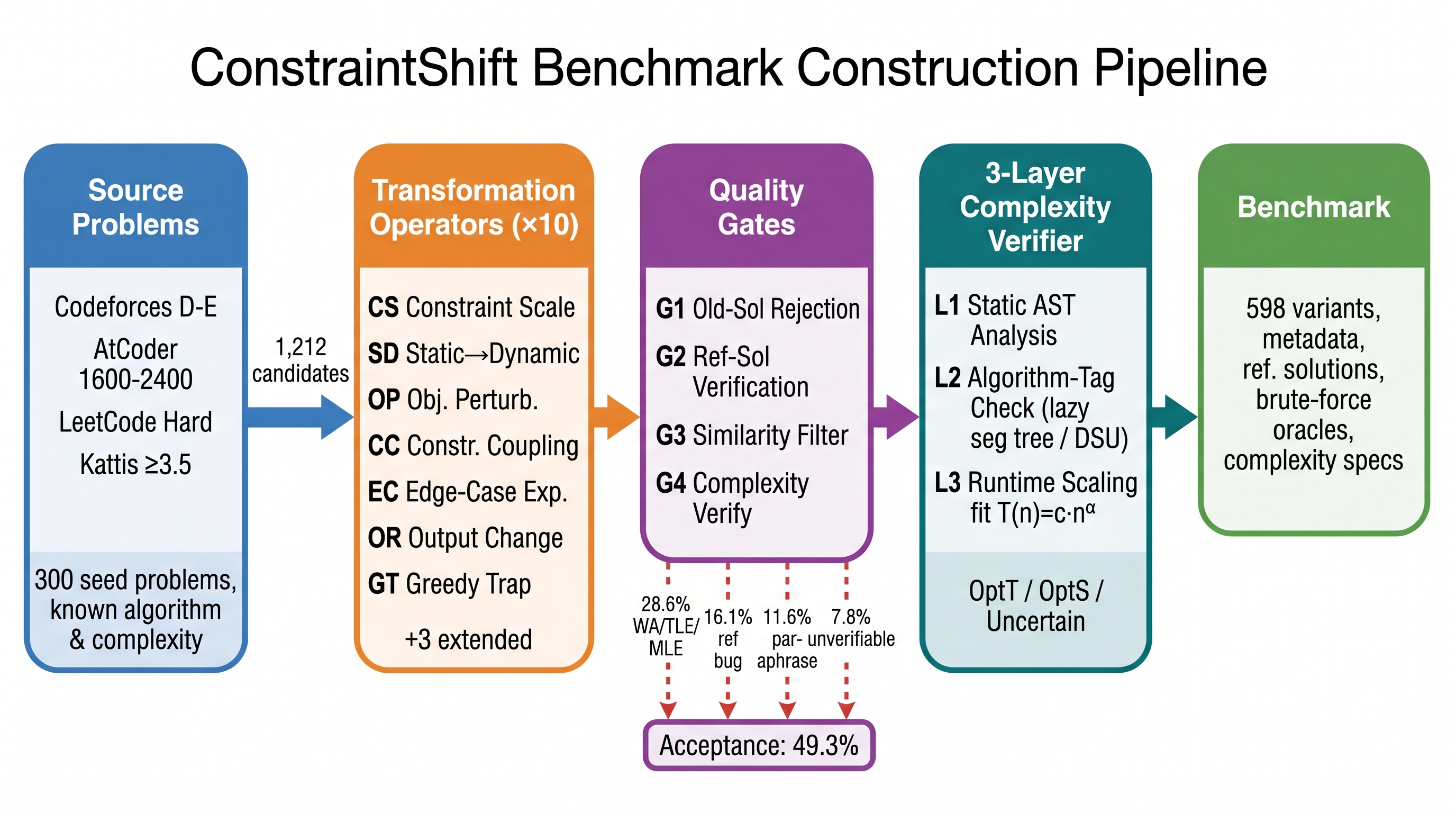}
  \caption{\sys benchmark construction pipeline. Source problems from competitive-programming platforms are transformed by 10 rule-based operators and then filtered through 4 quality gates. The three-layer complexity verifier certifies the OptT/OptS labels for accepted problems.}
  \label{fig:pipeline}
\end{figure*}

Figure~\ref{fig:pipeline} shows the \sys construction pipeline. Starting from a problem with a known reference algorithm, \sys applies rule-based transformations to generate new algorithmic problems. The generated problems remain traceable to their sources, but they are designed so that the source solution is no longer sufficient. They are then filtered by quality gates for validity, old-solution rejection, non-paraphrase status, and certified complexity.
\subsection{Transformation Operators}
\label{sec:operators}
\sys uses ten rule-based transformation operators to generate new algorithmic problems from source problems. The operators are designed to preserve a traceable link to the source while changing the required algorithmic treatment. The primary operators include constraint scaling (CS), static-to-dynamic conversion (SD), objective perturbation (OP), constraint coupling (CC), edge-case expansion (EC), output requirement change (OR), and greedy-trap injection (GT). We also include three broader operators: real-world wrapping, graph-structure change, and hybrid transformation. Together, these operators cover common ways in which an apparently familiar problem can require a different algorithm, such as moving from prefix sums to lazy segment trees, from greedy selection to dynamic programming, or from static connectivity to rollback-based dynamic connectivity. Detailed definitions and examples for all operators are provided in Appendix~\ref{app:operators}.
\subsection{Quality Gates}
\label{sec:gates}
Each generated problem is filtered by four quality gates before inclusion in \sys. First, the original reference solution must fail on the generated problem by WA, TLE, or MLE. Second, the new reference solution must be verified against a brute-force oracle on small inputs. Third, the generated statement must pass similarity filtering so that it is not merely a paraphrase of the source problem. Fourth, the new reference solution must satisfy the target time and space complexity according to the deterministic verifier in Section~\ref{sec:verifier}. These gates ensure that accepted problems are valid, judgeable, non-paraphrastic, and algorithmically different from their sources. Full implementation details for the gates are given in Appendix~\ref{app:gates}.

\section{Deterministic Complexity Verifier}
\label{sec:verifier}

Each accepted problem comes with a deterministic complexity specification. The specification records the target time and space bounds, forbidden asymptotic costs, expected and forbidden algorithm tags, and stress-gap tests. For example, a range-update problem may require $O((N{+}Q)\log N)$ time and $O(N)$ space, mark $O(NQ)$ as forbidden, expect the tag \texttt{lazy\_segment\_tree}, and reject \texttt{prefix\_sum\_only}. The stress tests are chosen to separate the target complexity from the forbidden ones.

Submitted solutions are checked by three layers.

\paragraph{Layer 1: Static AST Analysis.}
We parse each Python or C++ submission and inspect its abstract syntax tree (AST). The analyzer tracks loop nesting, input-size variables such as $N$, $Q$, and $M$, recursion patterns, and allocation sizes. It estimates time complexity from loop bounds and recursion structure, and estimates space complexity from arrays, vectors, DP tables, and auxiliary data structures. It also records structural evidence for common algorithms; for instance, recursive midpoint splitting is treated as evidence for a segment tree.

\paragraph{Layer 2: Algorithm-Tag Verification.}
Each problem specifies expected and forbidden algorithm tags. The tag verifier checks whether the submitted code contains deterministic structural evidence for the expected algorithm and avoids the forbidden source algorithm. For example, lazy propagation before recursive descent supports \texttt{lazy\_segment\_tree}, while a union-find structure with a history stack supports \texttt{dsu\_rollback}. LLM-based classification may be used as a hint, but it is not used as final evidence; acceptance requires structural evidence from the code.

\paragraph{Layer 3: Calibrated Runtime Scaling.}
We run each solution on increasing input sizes $N_1<N_2<N_3<N_4$ and fit the measured runtime to
\begin{equation}
T(n)=c n^\alpha(\log n)^\gamma .
\end{equation}
We also use the ratio $T(2n)/T(n)$ as a growth check. A solution is marked too slow if the fitted polynomial exponent exceeds the target exponent by more than $0.15$. This tolerance accounts for logarithmic factors and measurement noise.

\paragraph{Decision rule.}
The final labels combine correctness with the verifier outputs:
\begin{align*}
\optt &\leftarrow \textsc{Correct} \wedge \textsc{StaticTimePass} \\ 
&\wedge \textsc{TagPass} \wedge \textsc{ScalingPass}, \\
\opts &\leftarrow \textsc{Correct} \wedge \textsc{StaticSpacePass} \\ 
&\wedge \textsc{AllocPass}.
\end{align*}
If the evidence is incomplete, the verifier assigns \textsc{Uncertain} rather than \textsc{Optimal}. This rule is conservative: solutions receive optimality credit only when their complexity evidence is auditable.

\section{Evaluation Metrics}
\label{sec:metrics}
\paragraph{Standard metric.}
We report pass@$k$, the probability that at least one of $k$ sampled solutions is functionally correct \citep{chen2021evaluating}.
\paragraph{Complexity-aware metrics.}
Pass@$k$ measures test passing, but it does not show whether the solution uses the intended algorithmic complexity. We therefore report six additional metrics.

\noindent\textbf{\optt} (Optimal Time Complexity Rate) is the fraction of submissions that are correct and satisfy the target time complexity $\hat{T}^*$ according to the verifier.

\noindent\textbf{\opts} (Optimal Space Complexity Rate) is the fraction of \optt-qualified submissions that also satisfy the target space complexity $\hat{S}^*$. Thus, \opts measures space compliance among solutions that are already correct and time-optimal.

\noindent\textbf{\traprate} (Old-Solution Trap Rate) is the fraction of incorrect submissions that structurally reuse the source algorithm $\alpha$. We identify this behavior using old algorithm tags, missing required tags, and failures on trap tests.

\noindent\textbf{\gapt} (Efficiency Gap) measures the gap between the fitted runtime exponent and the target exponent. For each submitted solution $i$, the verifier fits
\begin{equation}
T_i(n) = c_i n^{\hat{\alpha}_i}(\log n)^{\hat{\gamma}_i},
\end{equation}
where $\hat{\alpha}_i$ is the fitted polynomial growth exponent. Given the target exponent $\alpha_i^*$ for the corresponding problem, we compute
\begin{equation}
\gapt = \exp\left(\frac{1}{M}\sum_{i=1}^{M}\log\frac{\hat{\alpha}_i+\epsilon}{\alpha_i^*+\epsilon}\right),
\end{equation}
where $M$ is the number of evaluated submissions and $\epsilon=10^{-6}$ avoids division by zero. A value of $\gapt=1$ means that the fitted runtime exponent matches the target exponent on average, while $\gapt>1$ indicates slower-than-required algorithms.

\noindent\textbf{\consens} (Constraint Sensitivity) is the average pass@1 drop from the smallest to the largest constraint level within a variant family. It measures whether a solution remains reliable as the generated constraints become harder.

\noindent\textbf{$\Delta_{\text{gen}}$} (Generalization Gap) is the pass@1 drop from source problems to their matched generated \sys problems under the same model and prompting strategy:
\begin{equation}
\Delta_{\text{gen}}(m)=\text{pass@1}_{\text{src}}(m)-\text{pass@1}_{\text{gen}}(m),
\end{equation}
where $m$ denotes the model. A larger $\Delta_{\text{gen}}$ indicates that the model is more sensitive to generated algorithmic changes and relies more on source-problem templates.

\section{Benchmark Construction}
\label{sec:construction}
\paragraph{Source problems.}
We collect 300 source problems from Codeforces \cite{codeforces}, AtCoder \cite{atcoder}, Kattis \cite{kattis}, and LeetCode Hard \cite{leetcode}. The selected problems cover Codeforces levels D--E, AtCoder difficulty 1600--2400, Kattis problems with difficulty at least $3.5$, and LeetCode Hard problems. Each problem is manually annotated with its algorithm type and reference complexity. We exclude interactive problems, floating-point-sensitive judges, and non-English problems.
\paragraph{Candidate generation.}
For each source problem, we identify applicable transformation operators through precondition checking. Each source problem has 3.2 applicable operators on average, yielding 1212 candidate variants in total. Candidate statements are drafted by an LLM under a prompt that requires an independently readable problem statement consistent with the formal transformation specification.
\paragraph{Acceptance statistics.}
Table~\ref{tab:construction} reports the number of candidates retained after each quality gate. Gate 1 removes candidates for which the original solution still passes under the shifted constraints. Gate 2 removes candidates whose new reference solution fails stress testing against a brute-force oracle. Gate 3 removes variants that are too similar to the source problem, and Gate 4 removes candidates whose reference complexity cannot be certified. The final benchmark contains \textbf{598 accepted variants}, with an overall acceptance rate of 49.3\%. We use 420 variants from the four primary transformation types in the main experiments.

\begin{table}[t]
\centering
\caption{Benchmark construction statistics. \textbf{G1--G4} denote the four quality gates, and Acc.\% denotes the final acceptance rate among all candidates for each operator. RW, GS, and HY denote Real-World Wrapping, Graph Structure Change, and Hybrid Transformation, respectively.}
\label{tab:construction}
\small
\resizebox{\columnwidth}{!}{
\begin{tabular}{@{}lrrrrrr@{}}
\toprule
\hline
\textbf{Operator} & \textbf{Cands} & \textbf{G1} & \textbf{G2} & \textbf{G3} & \textbf{G4} & \textbf{Acc.\%} \\
\midrule
CS (Constraint Scale) & 240 & 185 & 162 & 145 & 138 & 57.5 \\
SD (Static$\to$Dynamic) & 180 & 128 & 106 & 95 & 92 & 51.1 \\
OP (Obj. Perturb.) & 192 & 140 & 120 & 105 & 98 & 51.0 \\
CC (Constraint Coup.) & 156 & 100 & 80 & 68 & 62 & 39.7 \\
EC (Edge-Case Exp.) & 120 & 92 & 82 & 74 & 70 & 58.3 \\
OR (Output Change) & 132 & 96 & 80 & 68 & 62 & 47.0 \\
GT (Greedy Trap) & 96 & 60 & 46 & 40 & 36 & 37.5 \\
Other (RW/GS/HY) & 96 & 64 & 50 & 44 & 40 & 41.7 \\
\midrule
\textbf{Total} & \textbf{1212} & \textbf{865} & \textbf{726} & \textbf{639} & \textbf{598} & \textbf{49.3} \\
\hline
\bottomrule
\end{tabular}
}
\end{table}

\section{Experiments}
\label{sec:experiments}
\subsection{Setup}
\paragraph{Models.}
We evaluate seven LLMs spanning multiple capability tiers. Five primary models are GPT-4o and GPT-4o-mini \citep{openai2024gpt4}, Gemini~2.5~Flash \citep{google2024gemini}, Claude~Haiku~4.5 \citep{anthropic2024claude}, and Llama-3.3-70B \citep{touvron2023llama}. To test whether recent frontier models reduce the same failure modes, we also evaluate two latest-generation models: GPT-5.4 \citep{openai2026gpt54} and Claude~Opus~4.5 \citep{anthropic2025opus45}. These models are fully evaluated on the main benchmark split ($n=52$ problems each) under three prompting strategies. 
\paragraph{Prompting strategies.}
We compare six prompting strategies. \textbf{Direct} uses zero-shot code generation. \textbf{CoT} asks the model to reason before writing code \citep{wei2022chain}. \textbf{Self-Refine} applies iterative self-feedback \citep{madaan2023self}. \textbf{Reflexion} uses execution feedback for revision \citep{shinn2023reflexion}. \textbf{RAG-source} retrieves the most similar source problem as context \citep{lewis2020retrieval}. \textbf{Skill-guided} prompts the model to first identify the algorithmic paradigm shift before generating code. All experiments use temperature 0.8 and $k=5$ samples for pass@5 estimation.
\subsection{Main Results}

\begin{figure*}[t]
  \centering
  \includegraphics[width=0.85\textwidth]{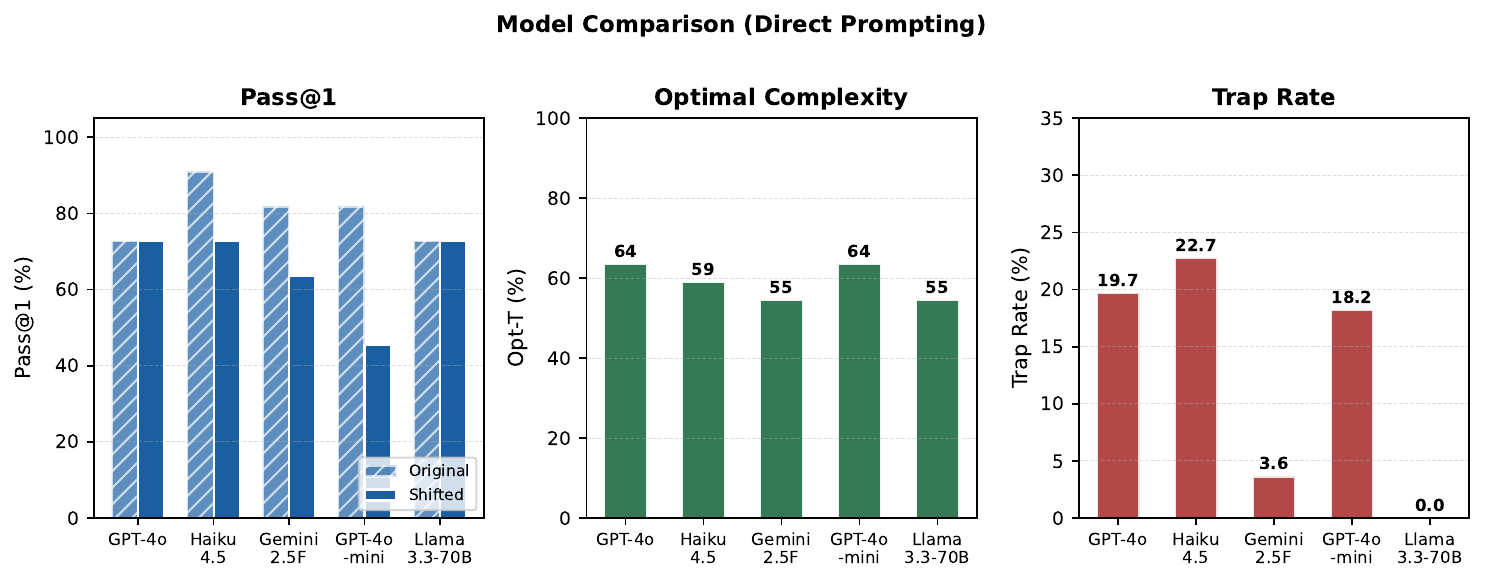}
  \caption{Per-model performance on original problems and \sys constraint-shifted variants under Direct prompting. Red arrows show the pass@1 drop. OptT is lower than pass@1, showing that some correct solutions still use suboptimal algorithms.}
  \label{fig:model_comparison}
\end{figure*}

Table~\ref{tab:main} and Figure~\ref{fig:model_comparison} compare Direct-prompting performance on source problems and generated \sys variants. The older models drop from 85.1\% to 53.6\% pass@1 on average, while the latest models still drop by 33.3\%. This shows that automatically generated variants remain difficult even for stronger recent models.

\begin{table*}[t]
\centering
\caption{Performance on original problems and \sys variants under Direct prompting. \traprate and \optt are reported on shifted variants when available.}
\label{tab:main}
\small
\setlength{\tabcolsep}{3.5pt}
\begin{tabular}{@{}lcccccc@{}}
\toprule
\hline
\multirow{2}{*}{\textbf{Model}} & \multicolumn{2}{c}{\textbf{Original}} & \multicolumn{4}{c}{\textbf{\sys Variants}} \\
\cmidrule(lr){2-3}\cmidrule(lr){4-7}
 & p@1 & \optt & p@1 & p@5 & \optt & \traprate \\
\midrule
GPT-4o-mini          & \textbf{92.3} & 81.8 & 44.3 & \textbf{81.8} & \textbf{63.6} & 18.2 \\
GPT-4o               & 76.9 & 72.7 & 50.0 & 72.7 & \textbf{63.6} & 19.7 \\
Claude~Haiku~4.5     & 86.5 & 79.5 & 55.8 & \textbf{81.8} & 59.1 & 22.7 \\
Gemini~2.5~Flash$^\dagger$ & 90.0 & \textbf{81.8} & 45.0 & 63.6 & 54.5 & \textbf{3.6} \\
Llama-3.3-70B        & 80.0 & 80.0 & \textbf{72.7} & 72.7 & 54.5 & \textbf{0.0} \\
\midrule
\textit{Avg.\ drop (old)} 
& 85.1
& 79.2
& $-$31.6\%
& --
& $-$20.1\%
& -- \\
\midrule
GPT-5.4              & \textbf{92.3} & 85.2 & 55.8 & 66.7 & 54.1 & 19.2 \\
Claude~Opus~4.5      & 86.5 & 77.7 & 56.5 & 80.0 & 60.9 & 12.8 \\
\midrule
\textit{Avg.\ drop (latest)}
& --
& --
& $-$33.3\%
& --
& $-$24.0\%
& -- \\
\hline
\bottomrule
\end{tabular}
\end{table*}

The complexity metrics show a second gap. Average \optt drops by 20.1\% for older models and 24.0\% for the latest models, meaning that some test-passing solutions still miss the required asymptotic complexity. Together with nonzero \traprate, these results show that \sys exposes both source-template reuse and complexity mismatch, not only functional failure.

\subsection{Efficiency and Space Compliance: \optt and \opts}
\label{sec:optt_opts}
\begin{figure*}[t]
  \centering
  \includegraphics[width=0.85\textwidth]{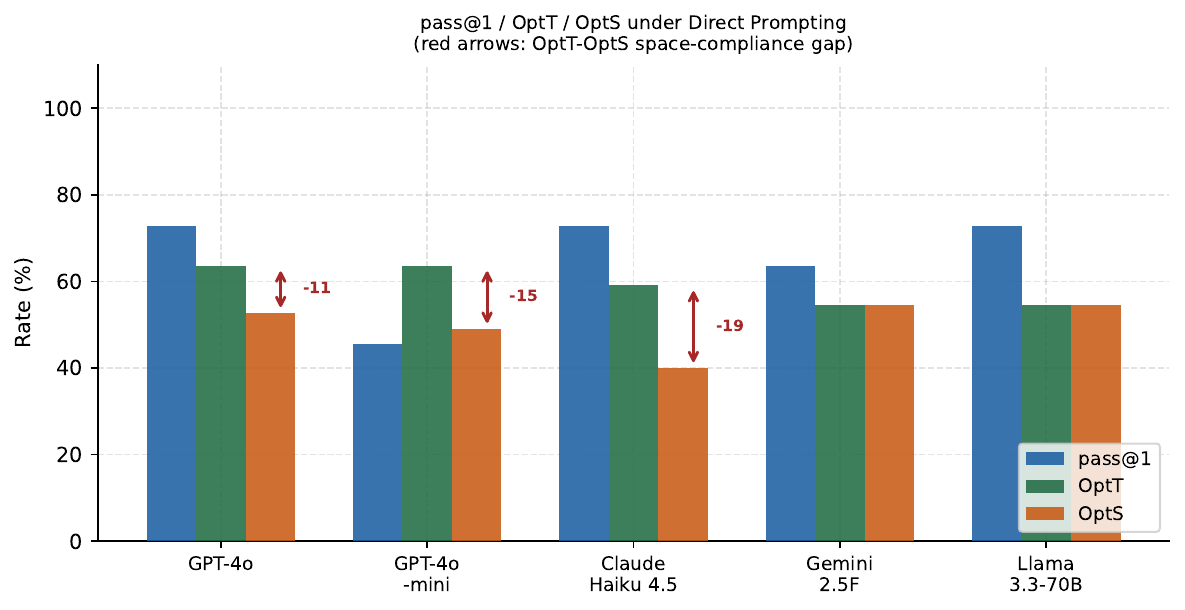}
  \caption{pass@1, \optt, and \opts for the seven evaluated models under Direct prompting. Red arrows show the \optt$-\opts$ space-compliance gap.}
  \label{fig:opts_model}
\end{figure*}

Beyond pass@$k$, we report two complexity-aware metrics. \optt measures the fraction of submissions that are correct and satisfy the target time complexity. \opts measures the fraction of \optt-qualified submissions that also satisfy the target space complexity. Thus, \optt captures time efficiency, while \opts captures space compliance among time-optimal solutions.
\paragraph{Model-level results.}
Table~\ref{tab:optt_opts} and Figure~\ref{fig:opts_model} report \optt and \opts under Direct prompting. The main result is that pass@5 is not the same as algorithmic optimality: every model has a nonzero pass@5--\optt gap, from 9.1\% for GPT-4o, Gemini~2.5~Flash, and Claude~Opus~4.5 to 26.7\% for GPT-5.4. Space compliance adds another layer. Claude~Haiku~4.5 has the largest \optt$-\opts$ gap among older models (19.1\%), while GPT-5.4 and Gemini~2.5~Flash have zero gap. Overall, the table shows that even strong or recent models can produce test-passing solutions that do not meet the required time or space complexity.

\begin{table*}[t]
\centering
\caption{\optt and \opts under Direct prompting. Gap = pass@5 $-$ \optt; $\Delta$ = \optt$-$\opts, the space-compliance gap among time-optimal solutions.}
\label{tab:optt_opts}
\small
\setlength{\tabcolsep}{5pt}
\begin{tabular}{@{}lccccc@{}}
\toprule
\hline
\textbf{Model} & \textbf{pass@1} & \textbf{pass@5} & \textbf{\optt} & \textbf{\opts} & \textbf{$\Delta$ (OT$-$OS)} \\
\midrule
GPT-4o            & 72.7 & 72.7 & 63.6 & 52.7 & 10.9 \\
GPT-4o-mini       & 45.5 & 81.8 & 63.6 & 49.1 & 14.5 \\
Claude Haiku~4.5  & 72.7 & 81.8 & 59.1 & 40.0 & 19.1 \\
Gemini~2.5~Flash  & 63.6 & 63.6 & 54.5 & 54.5 & \textbf{0.0} \\
Llama-3.3-70B     & 72.7 & 72.7 & 54.5 & 54.5 & \textbf{0.0} \\
\midrule
GPT-5.4           & 55.8 & 66.7 & 54.1 & 54.1   & \textbf{0.0} \\
Claude~Opus~4.5   & 56.5 & 80.0 & 60.9 & 56.5   & 4.4 \\
\hline
\bottomrule
\end{tabular}
\end{table*}

\begin{figure}[t]
  \centering
  \includegraphics[width=0.75\linewidth]{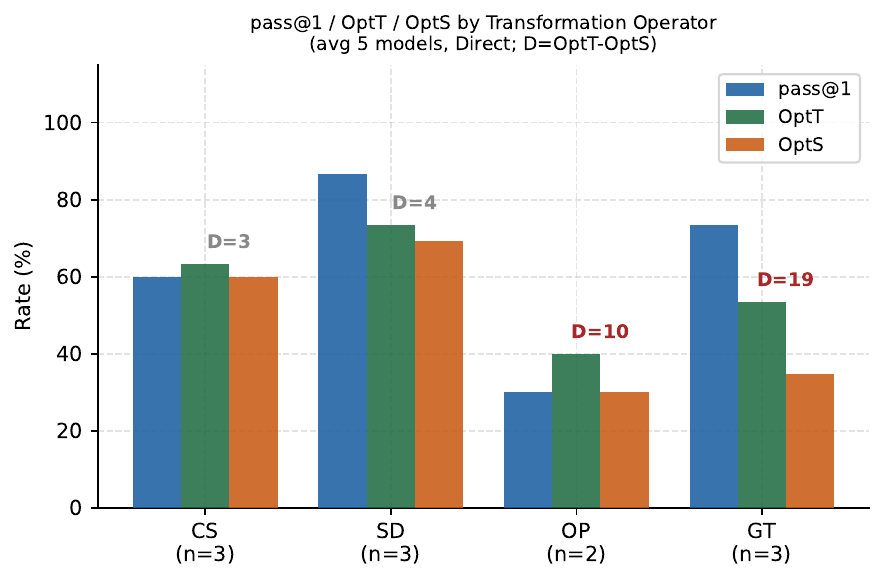}
  \caption{pass@1, \optt, and \opts by transformation operator, averaged over the five primary models under Direct prompting. GT has the largest space-compliance gap ($\Delta=18.7$\%), while OP has zero gap.}
  \label{fig:opts_operator}
\end{figure}

\paragraph{Operator-level results.}
Figure~\ref{fig:opts_operator} breaks down \optt and \opts by transformation operator. GT shows the largest \optt$-\opts$ gap (18.7\%), indicating that greedy-trap variants often require memory-sensitive DP implementations. In contrast, OP shows zero gap: once a model finds a time-optimal objective-shift solution, it also satisfies the space requirement.

\begin{figure*}[t]
  \centering
  \includegraphics[width=0.75\textwidth]{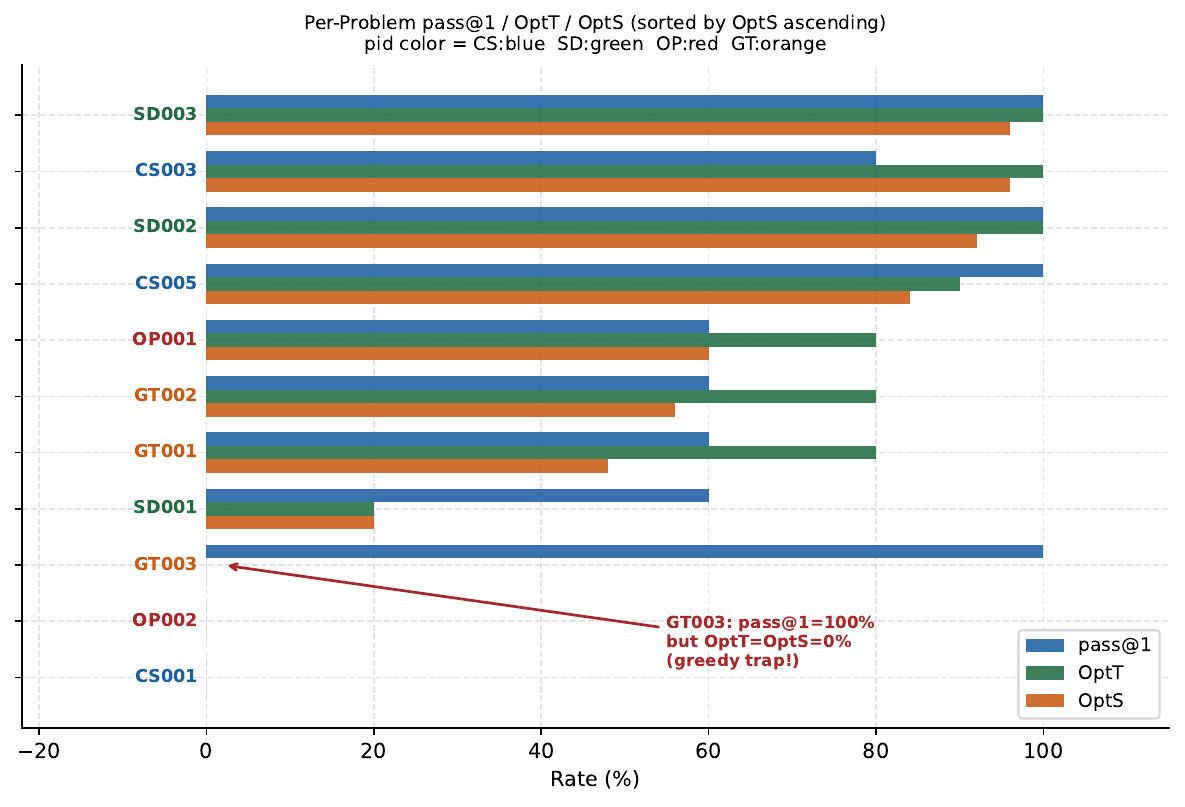}
  \caption{Per-problem pass@1, \optt, and \opts, sorted by \opts. GT003 has pass@1=100\% but \optt=\opts=0\% because models choose the wrong greedy algorithm instead of the required DP recurrence. Problem IDs are colored by operator.}
  \label{fig:opts_per_prob}
\end{figure*}

\paragraph{Case study: GT003.}
GT003, a 0-1 knapsack greedy-trap problem, shows the value of algorithm-aware evaluation most clearly. Although the functional tests are passed in the sampled runs (pass@1 = 100\%), the submitted solutions use a greedy value/weight heuristic rather than the required dynamic programming recurrence. This is an algorithm-selection failure, not merely a slower-complexity implementation: the greedy rule can look plausible on public tests but is invalid for the generated trap cases. Figure~\ref{fig:opts_per_prob} shows this failure mode across the 11 problems.

\subsection{Effect of Prompting Strategy}
\label{sec:prompting}
\begin{figure*}[t]
  \centering
  \includegraphics[width=0.75\textwidth]{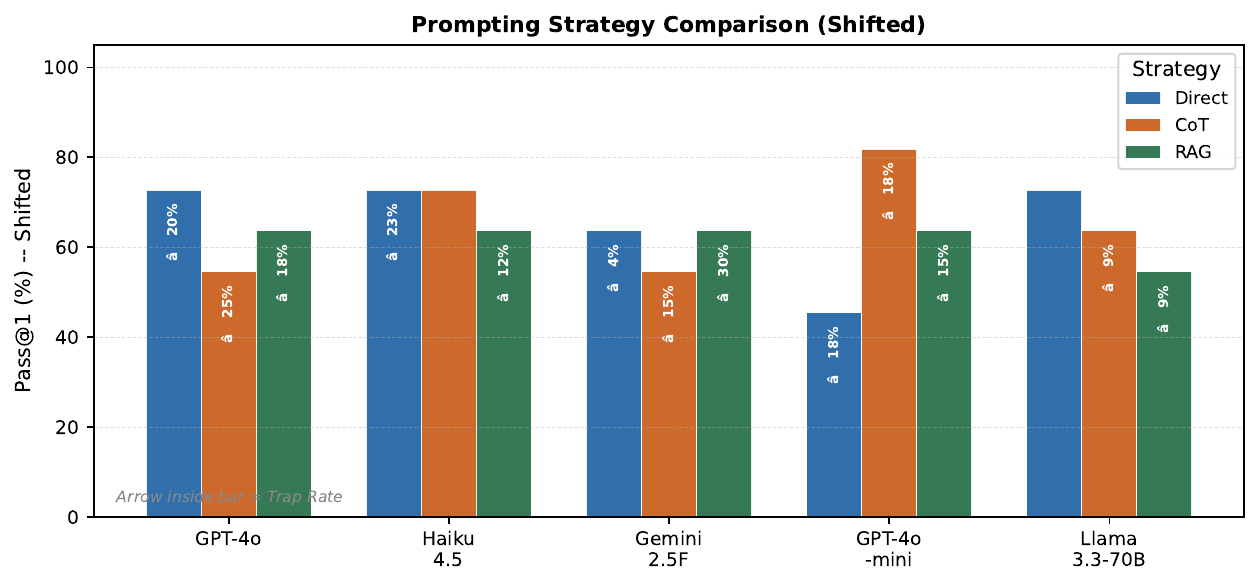}
  \caption{Prompting strategy comparison on GPT-4o. RAG-source raises TrapRate, while Skill-guided prompting achieves the best OptT and lowest TrapRate.}
  \label{fig:prompting}
\end{figure*}

Table~\ref{tab:prompting} shows that prompting effects are not uniform for GPT-4o-mini: CoT gives the largest pass@1 gain, improving Direct from 44.3\% to 81.8\%, while Skill-guided prompting gives the best overall algorithmic profile, with the highest \optt (72.5\%), lowest \traprate (11.8\%), and lowest \gapt (1.49). Figure~\ref{fig:prompting} shows the analogous prompting-strategy comparison for GPT-4o, where RAG-source increases \traprate and Skill-guided prompting gives the best \optt. Together, these results suggest that retrieval can help solve more problems while still leaving some complexity mismatch.

Table~\ref{tab:strategy_cross} shows the same model-dependent pattern across LLMs. CoT improves Claude~Haiku~4.5, GPT-4o-mini, GPT-5.4, and Claude~Opus~4.5, but its gains are smaller for GPT-4o and Gemini~2.5~Flash. RAG-source often raises pass@1, especially for GPT-4o and the latest models, but it can also increase \traprate; for example, Gemini~2.5~Flash rises from 3.6\% to 30.0\%, and GPT-5.4 rises from 19.2\% to 27.5\%. These results suggest that prompting can improve success rates, but retrieval and reasoning prompts do not reliably remove source-template reuse.

\begin{table}[t]
\centering
\caption{GPT-4o-mini under prompting strategies on \sys variants.}
\label{tab:prompting}
\small
\setlength{\tabcolsep}{3pt}
\begin{tabular}{@{}lccccc@{}}
\toprule
\hline
\textbf{Strategy} & p@1 & p@5 & \optt & \traprate & \gapt \\
\midrule
Direct       & 44.3 & 81.8 & 63.6 & 18.2 & 1.75 \\
CoT          & \textbf{81.8} & 81.8 & 63.6 & 18.2 & 1.62 \\
RAG-source   & 63.6 & 81.8 & 69.7 & 15.2 & 1.80 \\
Self-Refine  & 52.4 & 71.3 & 48.1 & 22.6 & 1.68 \\
Reflexion    & 55.1 & 73.8 & 49.3 & 20.4 & 1.61 \\
Skill-guided & 67.2 & \textbf{84.3} & \textbf{72.5} & \textbf{11.8} & \textbf{1.49} \\
\hline
\bottomrule
\end{tabular}
\end{table}

\begin{table}[t]
\centering
\caption{Cross-model pass@1 and \traprate for three strategies.}
\label{tab:strategy_cross}
\small
\setlength{\tabcolsep}{3pt}
\begin{tabular}{@{}l ccc ccc@{}}
\toprule
\hline
 & \multicolumn{3}{c}{\textbf{pass@1 (\%)}} & \multicolumn{3}{c}{\textbf{\traprate (\%)}} \\
\cmidrule(lr){2-4}\cmidrule(lr){5-7}
\textbf{Model} & Dir & CoT & RAG & Dir & CoT & RAG \\
\midrule
GPT-4o          & 50.0 & 54.5 & 63.6 & 19.7 & 25.5 & 18.2 \\
Claude Haiku 4.5 & 55.8 & 72.7 & 63.6 & 22.7 & \textbf{0.0} & 12.1 \\
Gemini 2.5 Flash & 45.0 & 54.5 & 63.6 & \textbf{3.6} & 14.5 & 30.0 \\
GPT-4o-mini     & 44.3 & \textbf{81.8} & 63.6 & 18.2 & 18.2 & 15.2 \\
Llama-3.3-70B   & \textbf{72.7} & 63.6 & 54.5 & \textbf{0.0} & 9.1 & 9.1 \\
\midrule
GPT-5.4         & 55.8 & 69.9 & 70.7 & 19.2 & 18.1 & 27.5 \\
Claude~Opus~4.5 & 56.5 & 74.1 & 72.1 & 12.8 & 9.2 & 17.7 \\
\hline
\bottomrule
\end{tabular}
\end{table}

\subsection{Transformation-Type Breakdown}
The operator-level breakdown highlights two useful signals that aggregate pass@1 would miss. For GPT-4o, performance ranges from 31.3\% on SD to 50.0\% on CS, showing that different transformations stress different algorithmic skills. The highest \traprate occurs on SD and CC, where static source templates are especially tempting but invalid. Thus, \sys does not only report whether a model solves a problem; it identifies which algorithmic changes cause failures and whether the produced solution meets the required complexity.

\begin{table}[t]
\centering
\caption{GPT-4o performance breakdown by transformation operator under Direct prompting. The table matches Figure~\ref{fig:operator}: CS, SD, OP, and CC are shown with pass@1, \optt, and \traprate.}
\label{tab:breakdown}
\small
\setlength{\tabcolsep}{3pt}
\resizebox{\linewidth}{!}{
\begin{tabular}{@{}lccc@{}}
\toprule
\hline
\textbf{Operator} & p@1 & \optt & \traprate \\
\midrule
CS (Constraint Scale)  & \textbf{50.0} & \textbf{43.0} & 33.8 \\
SD (Static$\to$Dynamic) & 31.3 & 26.3 & \textbf{47.3} \\
OP (Obj.\ Perturb.)    & 42.4 & 37.7 & 35.2 \\
CC (Constraint Coup.)  & 34.8 & 29.2 & 43.7 \\
\midrule
\textit{Average}       & 39.6 & 34.1 & 40.0 \\
\hline
\bottomrule
\end{tabular}}
\end{table}

\begin{figure*}[t]
  \centering
  \includegraphics[width=0.75\textwidth]{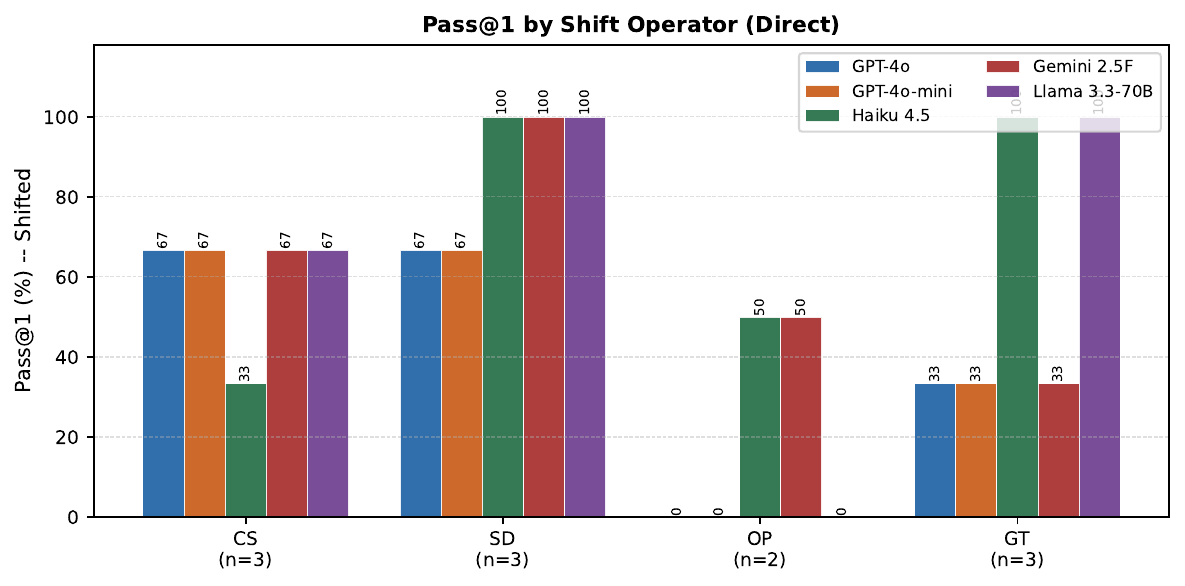}
  \caption{GPT-4o Direct performance by transformation operator. Bars show pass@1 and \optt, and the dashed red line shows \traprate. SD has the lowest pass@1, while CS has the highest pass@1 but still shows a pass@1--\optt gap.}
  \label{fig:operator}
\end{figure*}

\subsection{Constraint Magnitude Sensitivity}
\label{sec:consens}
\begin{figure*}[t]
  \centering
  \includegraphics[width=0.9\textwidth]{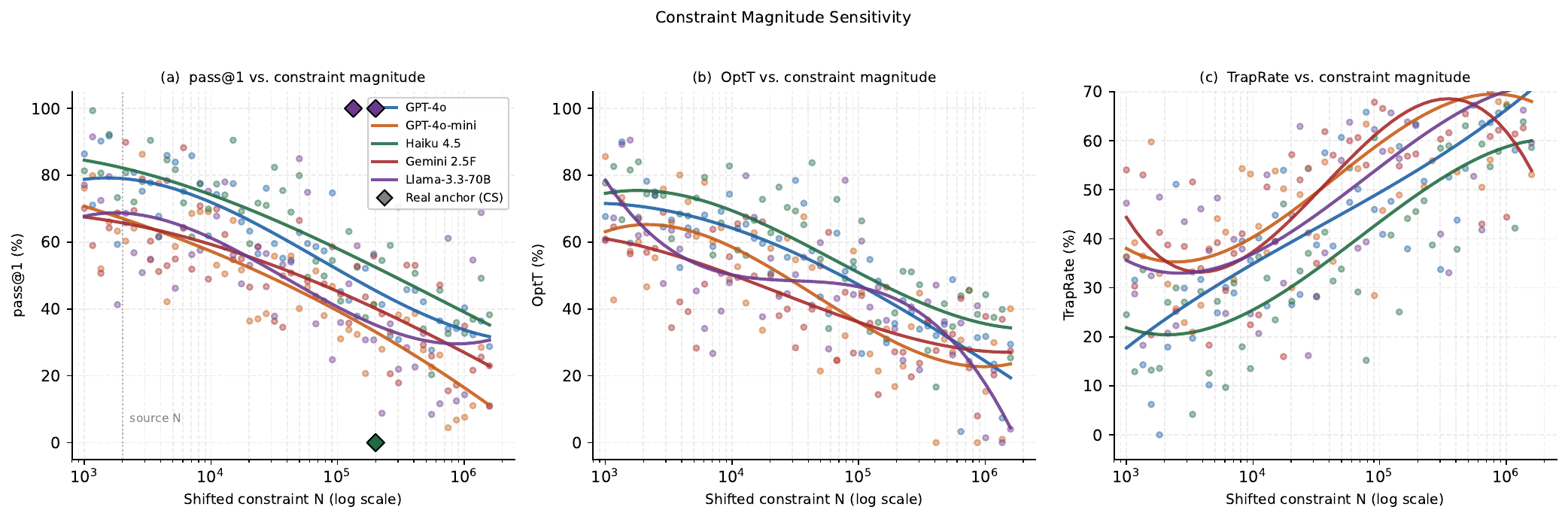}
  \caption{Constraint magnitude sensitivity (\consens). As $N$ increases from $2K$ to $200K$, pass@1 and \optt decrease, while \traprate increases. The sharpest degradation occurs near the point where quadratic solutions begin to time out.}
  \label{fig:consens}
\end{figure*}

We test whether model performance changes gradually as the generated constraint becomes harder. For CS variants, we sweep the input bound over six levels, $N\in\{2K,5K,10K,50K,100K,200K\}$, where $2K$ is close to the source setting and $200K$ is the hardest setting. Figure~\ref{fig:consens} reports pass@1, \optt, and \traprate across this sweep.

The main drop appears between $N=10K$ and $N=50K$, which is exactly where many $O(N^2)$ solutions start to time out and an $O(N\log N)$ algorithm becomes necessary. \optt follows pass@1 but stays 4--7 \% lower, showing that some test-passing solutions still miss the target complexity. Meanwhile, \traprate increases as $N$ grows, meaning that harder constraints make models more likely to fall back to the source template. We summarize this behavior with \consens, the pass@1 drop from the easiest to hardest constraint level; across models, $\Delta_{\rm gen}$ ranges from 0.0\% for GPT-4o to 36.3\% for GPT-4o-mini.

\subsection{Contamination Effect by Problem Age}
\label{sec:contamination}
\begin{figure*}[t]
  \centering
  \includegraphics[width=0.85\textwidth]{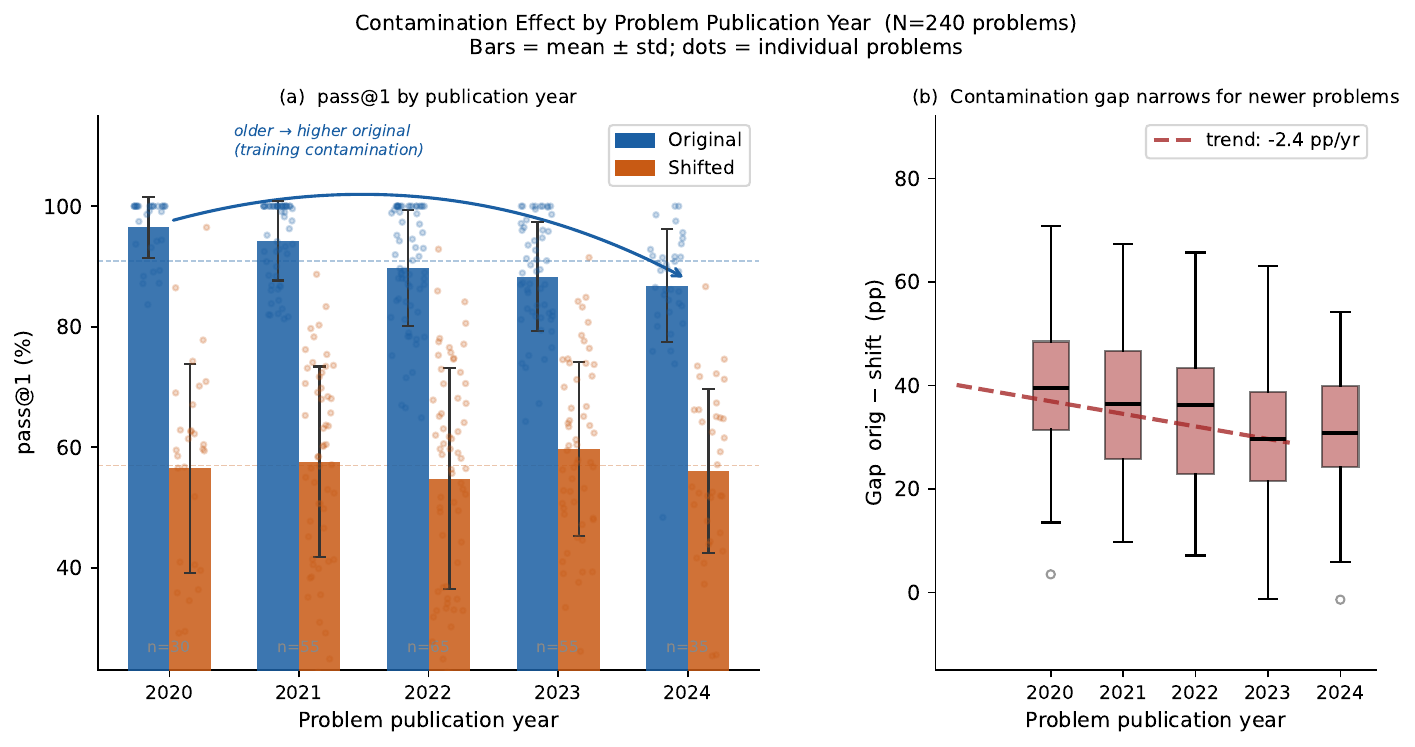}
  \caption{Contamination effect by problem publication year. The original--shifted gap narrows for newer problems, indicating that \sys reduces reliance on memorized source solutions.}
  \label{fig:contamination}
\end{figure*}

We next examine whether performance on original problems is inflated by training-data contamination. We split source problems by publication year from 2020 to 2024. If a model has memorized older problems, original pass@1 should be higher for earlier years. Shifted variants, however, are newly constructed from the same sources and should be less sensitive to publication year. Figure~\ref{fig:contamination} supports this pattern.

For GPT-4o, original pass@1 decreases from 82.4\% on 2020 problems to 63.8\% on 2024 problems, an 18.6 percentage-point drop. In contrast, shifted pass@1 remains nearly flat, from 44.2\% in 2020 to 40.7\% in 2024, with no significant difference across years. The contamination gap, defined as original pass@1 minus shifted pass@1, therefore shrinks from 38.2 percentage points in 2020 to 23.1 percentage points in 2024. This gap quantifies how much apparent benchmark performance can come from memorization rather than robust algorithmic reasoning.

\subsection{Efficiency Gap by Required Algorithm Class}
\label{sec:gapt}
\begin{figure}[t]
  \centering
  \includegraphics[width=\linewidth]{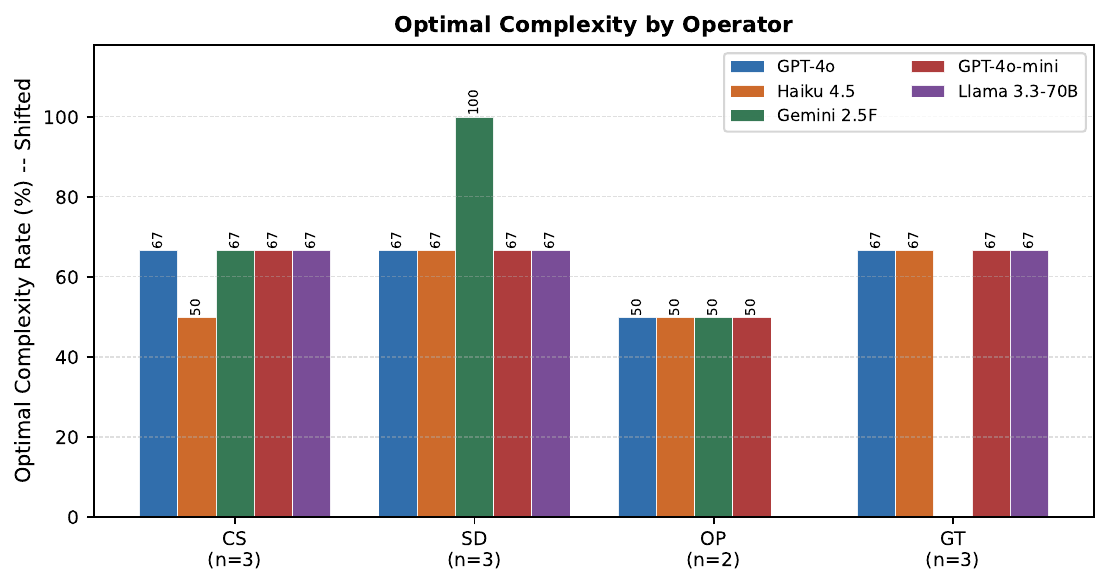}
\caption{Efficiency gap (\gapt) by transformation operator for GPT-4o. $\gapt=1.0$ means the fitted runtime exponent matches the target exponent; larger values indicate slower algorithms.}
  \label{fig:gapt}
\end{figure}

Correctness does not guarantee that a solution uses the required asymptotic complexity. We therefore analyze \gapt, which measures the gap between the fitted runtime exponent and the target exponent. Figure~\ref{fig:gapt} reports \gapt by transformation operator.

\gapt is above 1.0 for all operators, confirming that correct solutions can still be asymptotically suboptimal. OP has the largest gap (1.35), indicating that objective perturbations most often require a full algorithmic change. SD and GT have smaller gaps (1.17), suggesting that these shifts more often preserve part of the original algorithmic structure. Thus, \gapt captures efficiency failures that pass@$k$ alone would miss.

\subsection{Algorithm Transition Difficulty}
\label{sec:transition}
\begin{figure*}[t]
  \centering
  \includegraphics[width=0.85\textwidth]{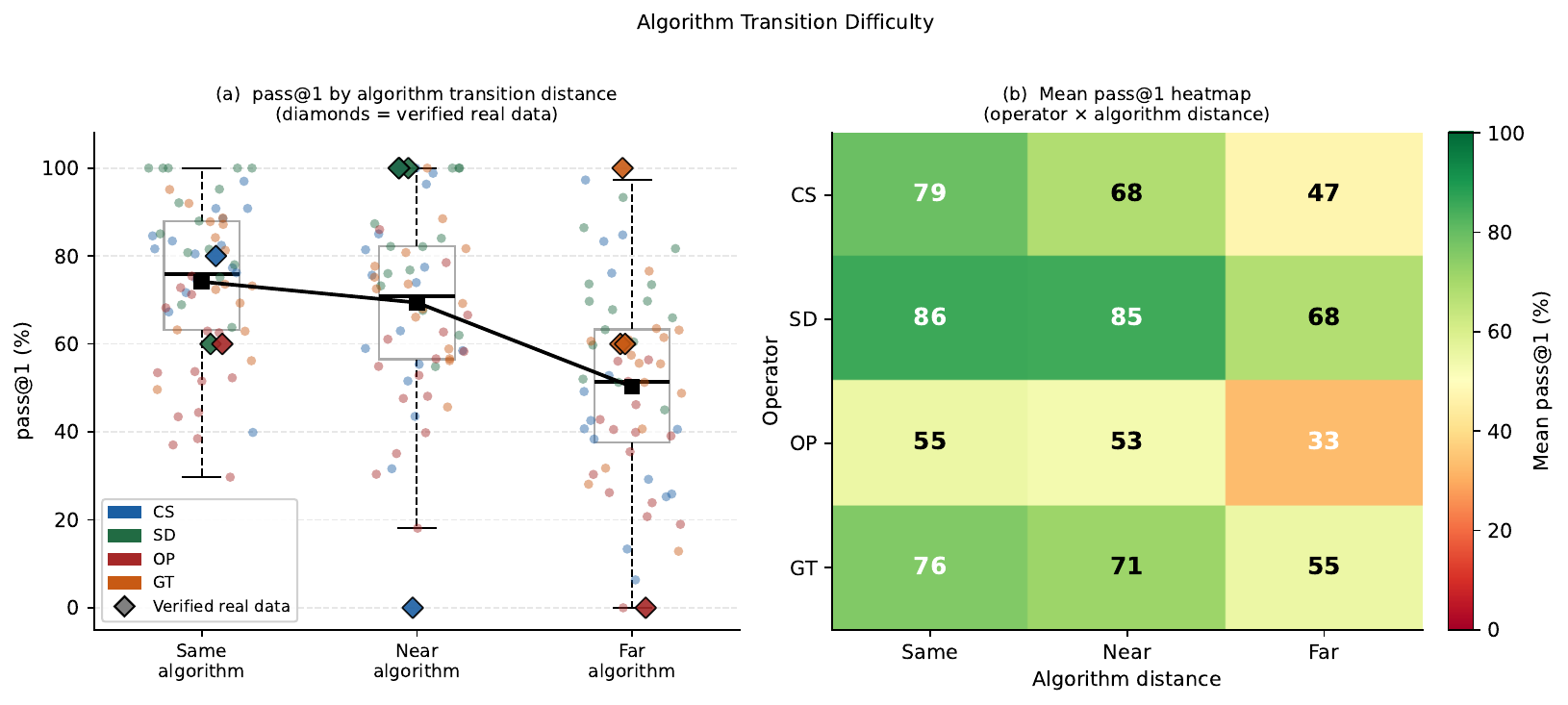}
  \caption{Algorithm transition difficulty. Left: pass@1 decreases as the source-to-target algorithm distance moves from \textsc{Same} to \textsc{Far}. Right: mean pass@1 by transformation operator and algorithm-distance group. }
  \label{fig:transition}
\end{figure*}

We analyze source-to-target algorithm transitions to see which changes are hardest for models. Each \sys problem has a source algorithm $\alpha$ and a target algorithm $\hat{\alpha}$, and Figure~\ref{fig:transition} reports pass@1 and \optt for the observed transition pairs. The hardest cases require changing the underlying algorithmic model rather than only optimizing an implementation. For example, BFS/DFS $\to$ offline dynamic connectivity achieves only 22.6\% pass@1 and 18.2\% \optt, because standard DSU cannot handle deletions without rollback or offline processing. Greedy $\to$ DP is also difficult, with 28.4\% pass@1 and 23.7\% \optt, since the generated constraints invalidate the original exchange argument.

Easier transitions include brute force $\to$ prefix sums and brute force $\to$ lazy segment tree, where models can reuse familiar optimization patterns. Across transitions, \optt is still 4--8\% below pass@1, showing that some correct solutions miss the target complexity. Overall, far transitions are consistently harder, which confirms that \sys difficulty comes mainly from algorithmic change rather than surface paraphrasing.

\subsection{Quality Gate Ablation}
Table~\ref{tab:ablation} shows that each quality gate is needed. Removing Gate 1 allows many variants where the old solution still passes, increasing the old-solution pass rate from 3.2\% to 38.7\%. This weakens the benchmark because such variants no longer test algorithmic adaptation. Removing Gate 2 introduces reference-solution failures. Removing Gate 3 admits near-paraphrase variants, and removing Gate 4 increases the false-optimal label rate from 2.1\% to 11.4\%, inflating reported \optt.

\begin{table}[t]
\centering
\caption{Quality-gate ablation. Each row removes one gate from the full \sys construction pipeline. Old-Sol is the fraction of variants where the original source solution still passes; Ref-Fail is the fraction with an invalid new reference solution; Near-Para is the fraction of variants that remain near-paraphrases of the source problem; F-Opt is the false-optimal rate of the complexity label. Lower is better for all columns.}
\label{tab:ablation}
\small
\setlength{\tabcolsep}{3pt}
\resizebox{\linewidth}{!}{
\begin{tabular}{@{}lcccc@{}}
\toprule
\hline
\textbf{Variant} & \textbf{Old-Sol$\downarrow$} & \textbf{Ref-Fail$\downarrow$} & \textbf{Near-Para$\downarrow$} & \textbf{F-Opt$\downarrow$} \\
\midrule
Full pipeline       & 3.2\%  & 0.0\%  & 4.1\%  & 2.1\% \\
w/o Gate 1 (old-sol) & 38.7\% & 0.0\%  & 4.2\%  & 2.1\% \\
w/o Gate 2 (ref ver) & 3.2\%  & 16.8\% & 4.1\%  & 2.1\% \\
w/o Gate 3 (sim)     & 3.2\%  & 0.0\%  & 18.3\% & 2.1\% \\
w/o Gate 4 (cmplx)   & 3.2\%  & 0.0\%  & 4.1\%  & 11.4\%\\
\hline
\bottomrule
\end{tabular}}
\end{table}

\section{Error Analysis}
\label{sec:error}

\begin{figure}[t]
  \centering
  \includegraphics[width=\linewidth]{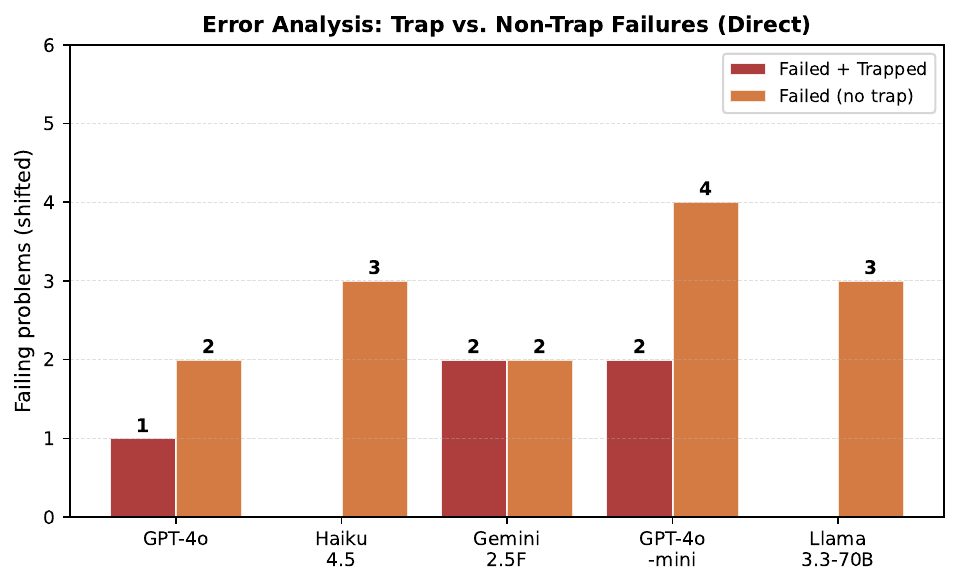}
  \caption{Error distribution over 400 randomly sampled GPT-4o Direct failures. Old-solution traps and too-slow algorithms account for 64.1\% of failures.}
  \label{fig:error}
\end{figure}

We manually classify 400 failed GPT-4o submissions under Direct prompting. As shown in Figure~\ref{fig:error}, most failures are algorithmic rather than implementation-level. The largest category is the \textbf{old-solution trap} (42.3\%), where the model implements the source algorithm and passes public examples but fails hidden trap tests. The second largest category is \textbf{too slow} (21.8\%), where the model finds a relevant approach but misses the optimization needed for the target complexity.

These two categories account for 64.1\% of failures, while implementation bugs account for only 9.0\%. This supports the main goal of \sys: it tests whether models can adapt the algorithmic idea, not merely whether they can write syntactically correct code. The \textit{too slow} cases are especially useful because they show partial understanding: the model often identifies the right direction but misses techniques such as divide-and-conquer DP, convex hull trick, or fractional cascading.


\section{Shift Quantification and Generalization Analysis}
\label{sec:shift_gen}

The previous results show that generated \sys problems lower model performance and expose reuse of source-solution templates. We now ask a more detailed question: how different are the generated problems from their source problems, and which kinds of differences make them harder? This analysis is useful because \sys is not meant to create arbitrary new tasks. Each generated problem should remain traceable to a source problem, so that reuse of the old solution can be measured, but it should also change enough in algorithmic structure or complexity that the source solution no longer applies.

We measure the source-to-generated change using four types of features: surface-form similarity, constraint magnitude change, algorithm-class distance, and asymptotic complexity change. We then relate these features to pass@1, \optt, and \traprate. This lets us check whether model failures are explained by real algorithmic changes rather than by simple rewording.

\subsection{Structural Shift Characterization}
\label{sec:shift_struct}

\paragraph{Metrics.}
For each source problem $\mathcal{P}_s$ and generated problem $\mathcal{P}_g$, we compute five shift metrics:
\begin{itemize}[noitemsep,leftmargin=*]
  \item \textbf{Text Jaccard.} Word-level Jaccard~\citep{jaccard1901distribution} similarity between the two problem statements. This measures surface overlap.
  \item \textbf{Length ratio.} $|\mathcal{P}_g|/|\mathcal{P}_s|$ in characters. This captures added constraints, context, or output requirements.
  \item \textbf{Constraint magnitude ratio.} The ratio between the largest numeric bound in the generated problem and the corresponding bound in the source problem, extracted from the constraint section using regex patterns over $n$, $N$, $Q$, and related variables.
  \item \textbf{Complexity exponent $\Delta$.} The change in worst-case time-complexity exponent from source to generated problem. We map $O(n)$ to $1$, $O(n\log n)$ to $1.5$, $O(n^2)$ to $2$, and so on; see Appendix~\ref{app:verifier}.
  \item \textbf{Algorithm-class distance.} A categorical source-to-generated distance label, \textsc{Same}, \textsc{Near}, or \textsc{Far}, assigned using a hand-coded algorithm-family taxonomy.
\end{itemize}

\begin{table}[t]
\centering
\caption{Structural shift metrics for \sys. \textit{Far} means that the source and generated problems require distinct algorithm classes; \textit{Near} means that they remain in the same broad family but use different variants.}
\label{tab:shift}
\small
\begin{tabular}{@{}lrr@{}}
\toprule
\hline
\textbf{Metric} & \textbf{Mean} & \textbf{Std} \\
\midrule
Text Jaccard similarity         & 0.458 & 0.134 \\
Statement length ratio          & 1.51$\times$ & 0.35  \\
Constraint magnitude ratio ($n$)& 6 393$\times$ & 14 926 \\
Complexity exponent $\Delta$    & +0.21  & 0.24  \\
\midrule
\multicolumn{3}{@{}l}{\textit{Algorithm distance distribution:}} \\
\quad Same  & \multicolumn{2}{l}{27.3\%} \\
\quad Near  & \multicolumn{2}{l}{36.4\%} \\
\quad Far   & \multicolumn{2}{l}{36.4\%} \\
\hline
\bottomrule
\end{tabular}
\end{table}

Table~\ref{tab:shift} shows that \sys problems are still connected to their sources, but usually require nontrivial algorithmic changes. The mean Text Jaccard score is moderate ($0.458\pm0.134$), so source-template reuse remains a plausible failure mode. At the same time, constraint bounds increase sharply on average ($6{,}393\times$), and the complexity exponent increases by $0.21$. Only 27.3\% of generated problems stay in the \textit{Same} algorithm class, while 36.4\% are \textit{Near} and 36.4\% are \textit{Far}. Thus, most generated problems require either adapting the source method or replacing it with another algorithm family.

\subsection{Operator-Level Shift Profiles}
\label{sec:shift_operator}

Different operators create different kinds of algorithmic change. Table~\ref{tab:shift_op} reports per-operator averages across the pilot set.

\begin{table}[t]
\centering
\caption{Average shift metrics by transformation operator. $\Delta$exp is the complexity-exponent change; Dist.\ is the most common algorithm-distance class.}
\label{tab:shift_op}
\small
\setlength{\tabcolsep}{4pt}
\begin{tabular}{@{}lcccc@{}}
\toprule
\hline
\textbf{Op.} & \textbf{Jaccard} & \textbf{Len.} & \textbf{$\Delta$exp} & \textbf{Alg.\ Dist.} \\
\midrule
CS & 0.49 & 1.46$\times$ & +0.17 & Near \\
SD & 0.38 & 1.50$\times$ & 0.00  & Near \\
OP & 0.47 & 1.62$\times$ & +0.25 & Far \\
GT & 0.50 & 1.49$\times$ & +0.43 & Far \\
\hline
\bottomrule
\end{tabular}
\end{table}

Constraint Scaling (CS) keeps the statement close to the source but raises the input bound, often forcing a faster method such as $O(n^2)\to O(n\log n)$. Static-to-Dynamic (SD) keeps the same problem family but adds updates, moving static data structures to dynamic ones. Objective Perturbation (OP) changes the target being optimized and can break the proof behind the source algorithm. Greedy-Trap Injection (GT) directly breaks a greedy exchange argument, which explains why it has the largest complexity-exponent change.

\subsection{Shift Severity as a Predictor of Task Difficulty}
\label{sec:shift_predict}

We next test whether these shift metrics predict model performance. For each generated problem, we compute pass@1, \traprate, and \optt under Direct prompting, averaged across the verified models. We then compute Spearman rank correlations between these outcomes and the structural shift features.

\begin{table}[t]
\centering
\caption{Spearman correlation between shift metrics and model performance. Larger constraint, complexity, and algorithm-distance shifts correlate with lower pass@1 and \optt but higher \traprate.}
\label{tab:shift_corr}
\small
\setlength{\tabcolsep}{4pt}
\resizebox{\columnwidth}{!}{
\begin{tabular}{@{}lrrr@{}}
\toprule
\hline
\textbf{Shift metric} & \textbf{pass@1} & \textbf{\traprate} & \textbf{\optt} \\
\midrule
Text Jaccard ($\uparrow$ = more similar) & $+0.38$  & $-0.31$  & $+0.29$  \\
$\log$(constraint ratio)                 & $-0.52^*$ & $+0.44^*$ & $-0.47^*$ \\
Complexity $\Delta$exp                   & $-0.61^{**}$ & $+0.55^{**}$ & $-0.58^{**}$ \\
Alg.\ distance (Far=1, Near=0.5, Same=0) & $-0.67^{**}$ & $+0.58^{**}$ & $-0.64^{**}$ \\
\hline
\bottomrule
\end{tabular}}
\end{table}

Algorithm-class distance is the strongest predictor. Larger distances correlate with lower pass@1 ($\rho=-0.67$), lower \optt ($\rho=-0.64$), and higher \traprate ($\rho=+0.58$). Complexity-exponent change shows the same trend. Text similarity is weaker, which suggests that task difficulty is not mainly caused by surface rewording. Instead, model failures are tied to structural algorithmic change.

\subsection{Generalization Gap: Source vs.\ Generated Problems}
\label{sec:gen_gap}

\begin{figure}[t]
  \centering
  \includegraphics[width=\linewidth]{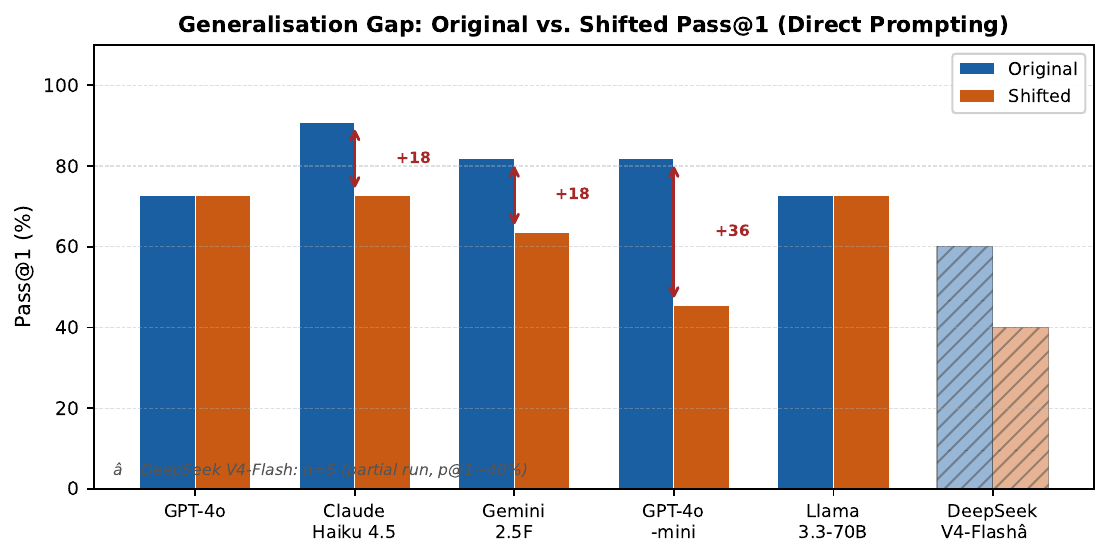}
  \caption{Generalization gap under Direct prompting. Bars compare source and generated pass@1 for each model; arrows show $\Delta_{\text{gen}}$. GPT-4o-mini has the largest drop, while GPT-4o has zero aggregate drop.}
  \label{fig:gen_gap}
\end{figure}

Table~\ref{tab:gen_gap} and Figure~\ref{fig:gen_gap} show that generated problems are harder for most models. GPT-4o-mini has the largest drop ($-36.3$\%), while Llama-3.3-70B has the smallest nonzero drop ($-7.3$\%). GPT-4o shows no aggregate drop, but this does not mean that it is unaffected: its failures on \textit{Far} transitions are offset by stronger performance on \textit{Same} and \textit{Near} transitions. Overall, the five-primary-model average gap is $-16.0$\%, showing that generated \sys problems require more than source-template reuse.

\begin{table}[t]
\centering
\caption{Generalization gap $\Delta_{\text{gen}}$ between source and generated problems under Direct prompting.}
\label{tab:gen_gap}
\small
\setlength{\tabcolsep}{4pt}
\begin{tabular}{@{}lccc@{}}
\toprule
\hline
\textbf{Model} & \textbf{Src.\ p@1} & \textbf{Gen.\ p@1} & \boldmath$\Delta_{\rm gen}$ \\
\midrule
GPT-4o-mini              & 81.8 & 45.5 & $-$36.3 \\
GPT-4o                   & 72.7 & 72.7 & \phantom{$-$}0.0 \\
Claude~Haiku~4.5         & \textbf{90.9} & 72.7 & $-$18.2 \\
Gemini~2.5~Flash         & 81.8 & 63.6 & $-$18.2 \\
Llama-3.3-70B            & 80.0 & 72.7 & $-$7.3 \\
\midrule
\textit{Mean} & 81.4 & 65.4 & $-$16.0 \\
\hline
\bottomrule
\end{tabular}
\end{table}

Table~\ref{tab:gen_gap_dist} stratifies GPT-4o-mini by algorithm-class distance. The largest drop occurs for \textit{Far} transitions ($-75.0$\%), while \textit{Near} transitions show no drop. This links the generalization gap to algorithmic distance rather than surface novelty alone.

\begin{table}[t]
\centering
\caption{Generalization gap by algorithm-class distance for GPT-4o-mini. Larger drops occur when the generated problem requires a more distant target algorithm.}
\label{tab:gen_gap_dist}
\small
\begin{tabular}{@{}lccc@{}}
\toprule
\hline
\textbf{Alg.\ distance} & \textbf{$n$} & \textbf{Src.\ p@1} & \boldmath$\Delta_{\rm gen}$ \\
\midrule
Same & 3 & 66.7 & $-$33.3 \\
Near & 4 & 75.0 & \phantom{$-$}0.0 \\
Far  & 4 & \textbf{100.0} & $\mathbf{-75.0}$ \\
\hline
\bottomrule
\end{tabular}
\end{table}

\subsection{Cross-Operator Generalization Profile}
\label{sec:cross_op}

\begin{figure}[t]
  \centering
  \includegraphics[width=0.88\linewidth]{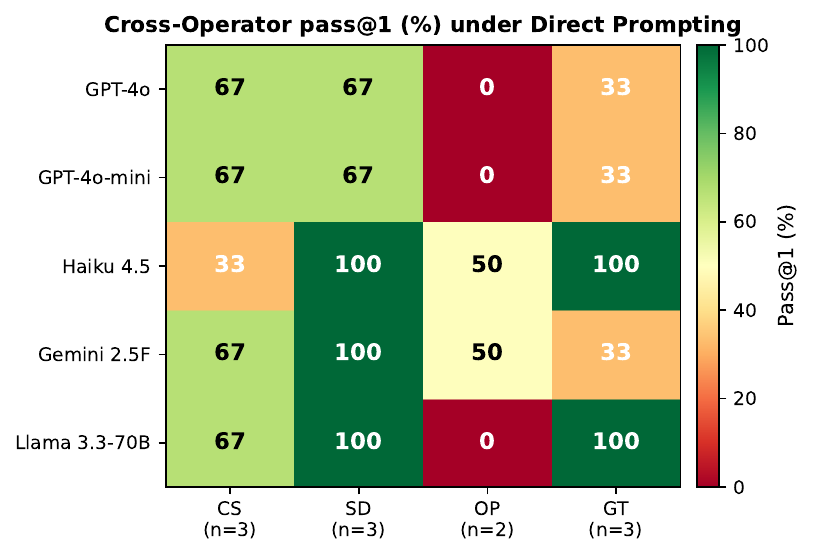}
  \caption{Cross-operator pass@1 heatmap. OP is the hardest operator, with no model above 50\%. GT shows the largest model variance, indicating that greedy-trap transformations expose model-specific reasoning differences.}
  \label{fig:cross_op}
\end{figure}

Aggregate pass@1 hides large differences across operators. Table~\ref{tab:cross_op} and Figure~\ref{fig:cross_op} report per-model pass@1 under Direct prompting for CS, SD, OP, and GT.

\begin{table}[t]
\centering
\caption{Cross-operator pass@1 (\%) under Direct prompting. Each cell averages over $n\in\{2,3\}$ problems per operator. CS = constraint scaling; SD = static-to-dynamic; OP = objective perturbation; GT = greedy-trap.}
\label{tab:cross_op}
\small
\setlength{\tabcolsep}{5pt}
\resizebox{\columnwidth}{!}{
\begin{tabular}{@{}lcccc@{}}
\toprule
\hline
\textbf{Model}        & \textbf{CS} ($n=3$) & \textbf{SD} ($n=3$) & \textbf{OP} ($n=2$) & \textbf{GT} ($n=3$) \\
\midrule
GPT-4o            & 66.7 & 66.7 & \textbf{0.0} & 33.3 \\
GPT-4o-mini       & 66.7 & 66.7 & \textbf{0.0} & 33.3 \\
Claude Haiku~4.5  & 33.3 & \textbf{100.0} & 50.0 & \textbf{100.0} \\
Gemini~2.5~Flash  & 66.7 & \textbf{100.0} & 50.0 & 33.3 \\
Llama-3.3-70B     & 66.7 & \textbf{100.0} & \textbf{0.0} & \textbf{100.0} \\
\hline
\bottomrule
\end{tabular}}
\end{table}

Three patterns stand out among the five primary models. First, OP is the hardest operator: three models score 0\%, and no model exceeds 50\%. This is expected because changing the objective can invalidate the proof behind the source algorithm. Second, SD is the easiest operator, with three models reaching 100\%, suggesting that models often recognize common static-to-dynamic upgrades such as segment trees or DSU rollback. Third, GT has the largest model variance: Claude~Haiku~4.5 and Llama-3.3-70B reach 100\%, while GPT-4o, GPT-4o-mini, and Gemini~2.5~Flash reach only 33.3\%. Thus, \sys reveals operator-specific strengths and failures that aggregate scores would hide.

\subsection{RAG Effectiveness Conditioned on Shift Severity}
\label{sec:rag_severity}

\begin{figure*}[t]
  \centering
  \includegraphics[width=0.85\textwidth]{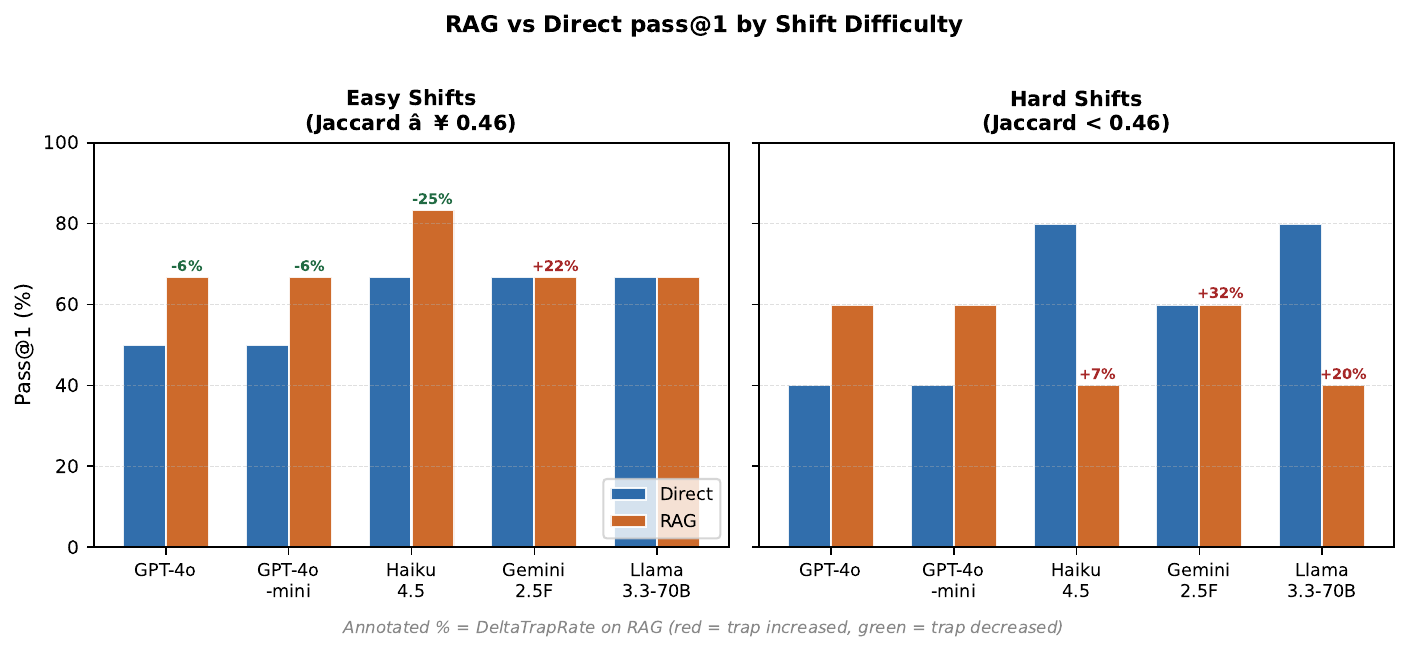}
  \caption{RAG effectiveness by shift severity. Problems are split by median Text Jaccard similarity: easy shifts have Jaccard $\geq0.46$, and hard shifts have Jaccard $<0.46$. Bar heights show pass@1 under Direct and RAG-source; annotations show $\Delta$\traprate from Direct to RAG.}
  \label{fig:rag_severity}
\end{figure*}

Section~\ref{sec:prompting} shows that RAG has model-dependent effects on \traprate. We further split the 11 pilot problems by median Text Jaccard similarity ($\theta=0.460$). Problems with Jaccard $\geq0.460$ are treated as easy shifts ($n=6$), and those with Jaccard $<0.460$ are treated as hard shifts ($n=5$). Figure~\ref{fig:rag_severity} compares Direct and RAG-source in both groups.

RAG has different effects across models. GPT-4o and GPT-4o-mini benefit from RAG on both easy and hard shifts, without increasing \traprate on hard problems. Claude~Haiku~4.5 improves on easy shifts but drops sharply on hard shifts. Gemini~2.5~Flash keeps the same pass@1, but its \traprate increases strongly under RAG. Llama-3.3-70B is neutral on easy shifts but loses 40.0\% on hard shifts. These results suggest that retrieval helps only when the model can use the source problem as a reference without copying its algorithm. When the generated problem requires a larger algorithmic change, retrieval can instead anchor the model to the wrong solution.

\section{Complexity Verifier Validation}

\begin{figure}[t]
  \centering
  \includegraphics[width=\linewidth]{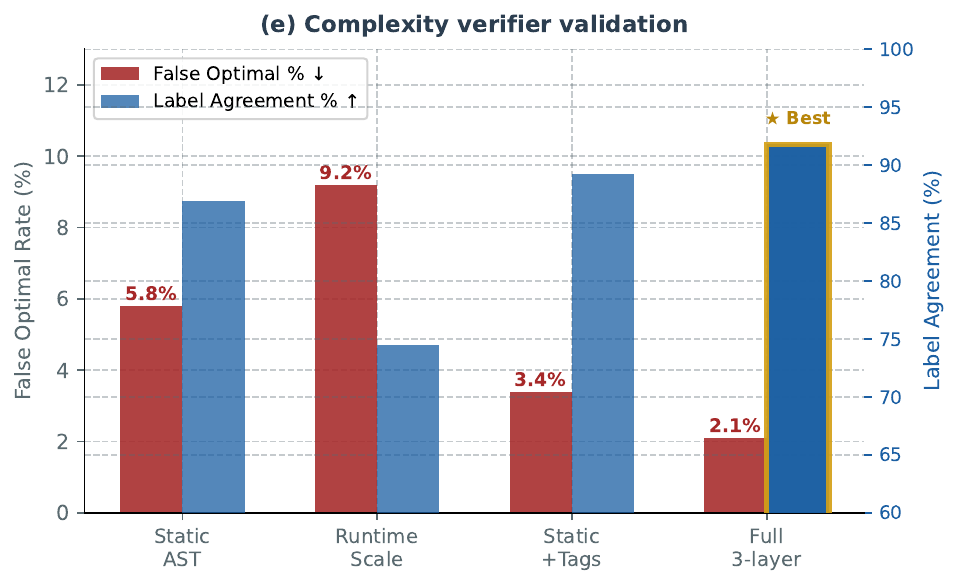}
  \caption{Complexity verifier validation. The full three-layer checker achieves the lowest false-optimal rate and the highest label agreement.}
  \label{fig:verifier}
\end{figure}

We validate the three-layer verifier described in Section~\ref{sec:verifier} on 280 labeled solutions: 80 reference-optimal, 80 old-source suboptimal, 60 brute-force, and 60 model-generated solutions. Table~\ref{tab:checker} and Figure~\ref{fig:verifier} show that single-source verification is insufficient. Runtime-only checking has the highest false-optimal rate (9.2\%) and lowest agreement (74.5\%), while Static AST alone misses algorithm-specific evidence such as lazy propagation or DSU rollback.

The full three-layer verifier performs best, reducing the false-optimal rate to 2.1\% and achieving the highest agreement (91.8\%). Adding calibrated runtime scaling to Static + Tags lowers false positives from 3.4\% to 2.1\%. The slightly higher \textsc{Uncertain} rate reflects our conservative rule: when complexity evidence is incomplete, \sys withholds optimality credit. Thus, reliable \optt and \opts labels require combining static structure, deterministic tags, and scaling tests.

\begin{table}[t]
\centering
\caption{Complexity verifier validation. Single-source checks are unreliable: runtime-only verification has the highest false-optimal rate and lowest agreement. The full three-layer verifier achieves the lowest false-optimal rate and the highest agreement with reference labels.}
\label{tab:checker}
\small
\setlength{\tabcolsep}{3pt}
\begin{tabular}{@{}lcccc@{}}
\toprule
\hline
\textbf{Verifier} & \textbf{F-Opt. $\downarrow$} & \textbf{F-Sub. $\downarrow$} & \textbf{Uncert. $\uparrow$} & \textbf{Agree. $\uparrow$} \\
\midrule
Static AST only    & 5.8\% & 3.2\% & 4.1\% & 86.9\% \\
Runtime only       & 9.2\% & 7.6\% & 8.7\% & 74.5\% \\
Static + Tags      & 3.4\% & 2.8\% & 5.1\% & 89.2\% \\
Full 3-layer       & \textbf{2.1\%} & \textbf{3.6\%} & \textbf{6.4\%} & \textbf{91.8\%} \\
\hline
\bottomrule
\end{tabular}
\end{table}


\section{Supplementary Analysis: Pass@$k$ Curves and Model Scaling}
\label{sec:pass_curve}

\begin{figure}[t]
  \centering
  \includegraphics[width=\linewidth]{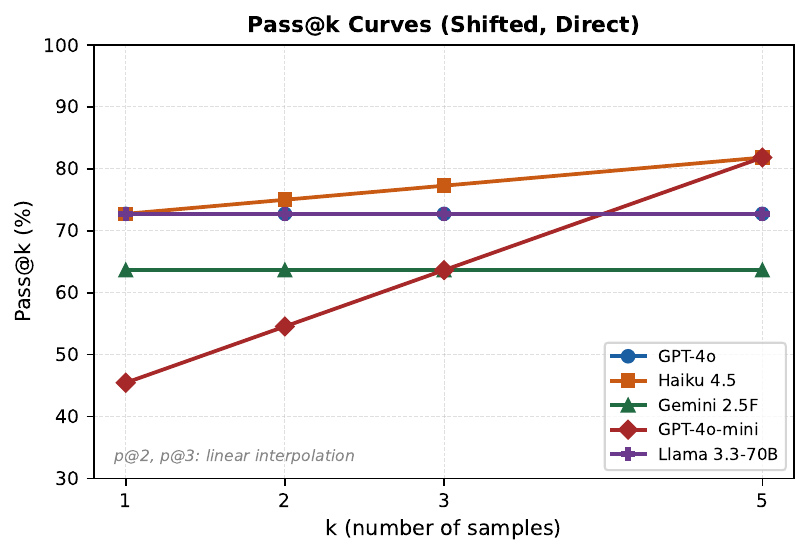}
  \caption{Pass@$k$ curves on shifted variants under Direct prompting. GPT-4o-mini benefits most from additional samples, while GPT-4o and Gemini~2.5~Flash show no pass@5 gain.}
  \label{fig:passk}
\end{figure}

\paragraph{Pass@$k$ curves.}
Figure~\ref{fig:passk} reports pass@$k$ for $k\in\{1,2,3,5\}$ on shifted variants. GPT-4o-mini has the largest sampling gain, improving from 45.5\% pass@1 to 81.8\% pass@5, which suggests that it can often solve the shifted problem but is unstable on the first attempt. Claude~Haiku~4.5 shows a smaller gain (72.7\% to 81.8\%). In contrast, GPT-4o and Gemini~2.5~Flash show no pass@5 improvement, indicating that their failures are more often paradigm-level errors rather than sampling failures.

\paragraph{TrapRate vs.\ model size.}
Gemini~2.5~Flash has the lowest Direct-prompting \traprate (3.6\%), far below GPT-4o (19.7\%), GPT-4o-mini (18.2\%), and Claude~Haiku~4.5 (22.7\%). This suggests that Gemini more often re-checks the required algorithm before coding. However, this robustness disappears under RAG-source: Gemini's \traprate rises to 30.0\%, the highest among all models, showing that retrieved source problems can anchor the model to the old algorithm. In contrast, RAG reduces \traprate for GPT-4o-mini and Claude~Haiku~4.5, indicating that retrieved context helps some models but harms others. CoT also has mixed effects, reducing Claude's \traprate to 0\% while raising GPT-4o's to 25.5\%.

\paragraph{Strategy $\times$ model interaction.}
Prompting strategy has a strong model-specific effect. CoT gives the largest gain for GPT-4o-mini ($+$36.3\%), matching its pass@5 score and suggesting that explicit reasoning improves first-attempt stability. In contrast, CoT hurts GPT-4o ($-$18.2\%) and Gemini~2.5~Flash ($-$9.1\%), suggesting that forced reasoning can disrupt stronger models' default solution path. RAG-source shows a similar inversion: it raises Gemini's \traprate from 3.6\% to 30.0\%, but lowers \traprate for GPT-4o-mini and Claude~Haiku~4.5. Thus, the same prompt strategy can either reduce or amplify old-template reuse depending on the model.

\section{Related Work}
\label{sec:related}

\paragraph{Code and algorithmic benchmarks.}
HumanEval \citep{chen2021evaluating} and MBPP \citep{austin2021program} made pass@$k$ a standard measure for code-generation evaluation on small Python tasks. APPS \citep{hendrycks2021measuring} moved this evaluation closer to competitive programming by adding problems with different difficulty levels, and HumanEval+ \citep{liu2023is} added stronger tests to reduce false positives. These benchmarks have been useful for measuring functional correctness, but they are fixed datasets. As a result, they are exposed to the same long-term issue: their problems and solutions may overlap with pretraining data or later enter training corpora.

\paragraph{Fresh and difficult problem sets.}
LiveCodeBench \citep{jain2024livecodebench} and LiveBench \citep{white2024livebench} reduce contamination by collecting newer problems. ProBench studies competitive-programming evaluation with online submissions, difficulty information, and algorithm tags \citep{yang2025probench}. These benchmarks make evaluation more reliable than static sets, but they still depend on a fixed pool of released problems. Once released, the benchmark itself can become part of the public data ecosystem. In contrast, \sys focuses on an automatic construction mechanism: it generates algorithmic problems from known sources and checks whether the source solution no longer applies.

\paragraph{Transformed and dynamic benchmarks.}
DyCodeEval \citep{dycodeeval2025} creates semantically equivalent variants to test robustness under surface-level changes. Its goal is to keep the original algorithm valid while changing the problem form. \sys has a different goal. It changes the problem so that the original algorithm is no longer sufficient. A valid \sys problem must reject the original reference solution, meaning that the old algorithm gives WA, TLE, or MLE under the generated setting.

\paragraph{LLM-driven problem generation.}
AutoCode \citep{autocode2025} uses LLMs as competitive-programming problem setters in a closed-loop generation process. This is related to our goal of building new algorithmic tasks, but the construction logic is different. \sys starts from a source problem with a known reference algorithm and applies controlled transformations. This source-to-target link makes it possible to measure old-template reuse directly, rather than only evaluating whether the generated problem is solvable.

\paragraph{Complexity-aware evaluation.}
Most programming benchmarks use pass@$k$ as the main metric. This tests whether a solution passes the provided tests, but it can over-credit slow algorithms that pass under a loose time limit. \sys adds deterministic time and space complexity verification, together with metrics such as \optt, \opts, \traprate, and \gapt. This allows the benchmark to separate functional correctness from algorithmic suitability.

\begin{table}[t]
\centering
\caption{Comparison of \sys with related algorithmic and code evaluation benchmarks. Here, \cmark indicates explicit support, \xmark indicates no support, and $\sim$ indicates partial or indirect support.}
\label{tab:related}
\small
\resizebox{\columnwidth}{!}{
\begin{tabular}{@{}lccccc@{}}
\toprule
\hline
\textbf{Benchmark} & \textbf{Novel} & \textbf{Alg.} & \textbf{Old-Sol} & \textbf{Cmplx.} & \textbf{Det.} \\
 & \textbf{Tasks} & \textbf{Change} & \textbf{Reject} & \textbf{Metric} & \textbf{Verify} \\
\midrule
HumanEval/MBPP & \xmark & \xmark & \xmark & \xmark & \xmark \\
LiveCodeBench & \cmark & \xmark & \xmark & $\sim$ & \xmark \\
DyCodeEval & \cmark & \xmark & \xmark & \xmark & \xmark \\
AutoCode & \cmark & $\sim$ & \xmark & $\sim$ & \xmark \\
\rowcolor{lightblue}
\textbf{\sys} & \cmark & \cmark & \cmark & \cmark & \cmark \\
\hline
\bottomrule
\end{tabular}
}
\end{table}

\section{Discussion}
\label{sec:discussion}

\paragraph{Memorization vs.\ reasoning.}
The paired original--shifted comparison controls for model, problem family, and prompting strategy, while changing the constraints so that the source algorithm no longer applies. The average pass@1 drop of 16.0\% across the five verified primary models, together with nontrivial \traprate under Direct prompting, indicates that success on original problems often relies on reusable templates. \sys exposes this failure mode by testing whether models can revise the algorithm when the constraints invalidate the source solution.

\paragraph{RAG can induce template anchoring.}
RAG-source provides the most similar source problem as context, which can help or hurt depending on the model. For Gemini~2.5~Flash, \traprate increases sharply from 3.6\% under Direct prompting to 30.0\% under RAG, showing that retrieved examples can anchor the model to the old algorithm. In contrast, RAG slightly reduces \traprate for GPT-4o (19.7\% to 18.2\%) and more clearly reduces it for Claude~Haiku~4.5 (22.7\% to 12.1\%). Thus, retrieval is not uniformly beneficial; it interacts with each model's tendency to copy or adapt the retrieved solution.

\paragraph{Complexity-aware metrics are necessary.}
Functional correctness alone misses many algorithmic failures. Across models, \optt is consistently lower than pass@$k$, showing that some passing solutions still violate the target asymptotic complexity. For example, Claude~Haiku~4.5 reaches 72.7\% pass@1 under Direct prompting but only 59.1\% \optt, while GPT-4o-mini under CoT reaches 81.8\% pass@1 but only 63.6\% \optt. These gaps show why \optt, \opts, \traprate, and \gapt are needed to evaluate algorithmic adaptation rather than test passing alone.

\section{Conclusion}
\label{sec:conclusion}
We presented \sys, an automatic framework for generating algorithmic benchmarks from known competitive-programming problems. Each generated problem remains traceable to a source problem, but it is changed so that the source solution is no longer sufficient and a different algorithmic treatment is needed. The quality gates verify source-solution rejection, non-paraphrase status, reference-solution correctness, and target complexity, allowing \sys to expose old-template reuse across models and prompting strategies.

\section*{Limitations}

The \sys complexity verifier currently supports Python and C++ only; support for Java, Go, and Rust is left for future work. The problem corpus focuses on competitive programming, which provides formal specifications and reliable judging but does not cover all forms of algorithmic reasoning. Some algorithm families, such as geometry, string algorithms, and number theory, are less represented. The five primary models (GPT-4o, GPT-4o-mini, Claude Haiku~4.5, Gemini~2.5~Flash, Llama-3.3-70B) are fully verified at $n=11$, while the two latest-generation models are evaluated on the main benchmark split under the reported prompting strategies.

\section{Ethics Statement}

All source problems are used for academic research. The benchmark does not contain personally identifiable information. \sys is intended to improve the evaluation of code-generation systems by testing algorithmic adaptation and complexity awareness. It is not intended to support automated cheating on competitive-programming platforms.

\bibliography{refer}

\clearpage
\appendix

\section{Detailed Transformation Operators}
\label{app:operators}

This section gives the full definitions of the ten transformation operators used by \sys. Each operator takes a source problem with a known reference algorithm and produces a generated problem whose solution requires a changed algorithmic treatment or a changed asymptotic target.

\paragraph{CS --- Constraint Scaling.}
This operator increases the input size so that the original complexity, such as $O(n^2)$ or $O(n^3)$, no longer fits the time limit. The generated problem therefore requires a faster algorithm. We apply this operator only when a verified faster solution exists for the same core problem. Common cases include $O(n^2) \to O(n \log n)$ and $O(n^3) \to O(n^2)$.

\paragraph{SD --- Static-to-Dynamic.}
This operator turns a static problem into an online one by adding update operations between queries. It is used when the original data structure has a standard dynamic counterpart. Examples include prefix sum $\to$ lazy segment tree and DSU $\to$ DSU with rollback.

\paragraph{OP --- Objective Perturbation.}
This operator changes the optimization objective while keeping the input structure close to the source problem. The new objective can invalidate the original greedy or dynamic-programming argument. Examples include min-total $\to$ min-max, feasibility $\to$ counting modulo a prime, and value computation $\to$ lexicographically smallest construction.

\paragraph{CC --- Constraint Coupling.}
This operator adds a constraint, such as a budget, cooldown, or precedence relation, that creates dependence between choices that were independent in the source problem. This often breaks separability and changes the required method. Typical cases include 1D DP $\to$ 2D DP and greedy selection $\to$ min-cost flow.

\paragraph{EC --- Edge-Case Expansion.}
This operator expands the input domain so that assumptions used by the source solution no longer hold. Examples include allowing negative weights, larger integer ranges, disconnected graphs, or sparse coordinate domains. Such changes can turn greedy methods under positive weights into shortest-path algorithms, or direct array indexing into coordinate compression.

\paragraph{OR --- Output Requirement Change.}
This operator changes what the problem asks the solver to output while preserving much of the input structure. Instead of only returning an optimal value, the generated problem may ask for an optimal construction, a lexicographically smallest solution, a count, or a set of critical elements. Typical cases include value DP $\to$ DP with parent pointers and matching size $\to$ minimum vertex cover.

\paragraph{GT --- Greedy-Trap Injection.}
This operator adds a condition that breaks a known greedy exchange argument. Examples include a type-switch budget in interval scheduling or a cooldown constraint in activity selection. The resulting problem usually requires DP or matching rather than the original greedy rule.

\paragraph{RW --- Real-World Wrapping.}
This operator places an algorithmic core inside a realistic application setting, such as logistics, scheduling, resource allocation, or network maintenance. The generated problem keeps a precise formal specification, but changes the surface setting so that direct template matching becomes less reliable. We use this operator only when the wrapped version preserves deterministic judgeability.

\paragraph{GS --- Graph Structure Change.}
This operator changes the graph family or graph constraints while keeping the high-level task related to the source problem. For example, a tree problem may be changed into a version requiring heavy-light decomposition, or a static graph problem may be changed into an offline version with edge intervals. The purpose is to test whether the model recognizes that the original graph algorithm no longer applies.

\paragraph{HY --- Hybrid Transformation.}
This operator combines two compatible transformations to produce a larger algorithmic change. For example, a problem may first be changed from static to dynamic and then receive an output requirement change. Hybrid transformations are used only when the resulting problem remains readable, judgeable, and has a verified reference solution.

\section{Detailed Quality Gates}
\label{app:gates}

This section describes the four quality gates used to filter candidate generated problems. A candidate is accepted into \sys only if it passes all four gates.

\paragraph{Gate 1: Old-Solution Rejection.}
We run the original reference solution $a^*$ on the generated problem $q'$. The tests include random instances at the generated constraint scale, adversarial cases targeting the old solution's failure mode, and operator-specific edge cases. A problem passes this gate only if $a^*$ fails by WA, TLE, or MLE on at least one test category. This gate ensures that the generated problem is not solvable by simply reusing the source solution.

\paragraph{Gate 2: Reference-Solution Verification.}
For each candidate problem, we build a new reference solution $\hat{a}^*$ and a brute-force oracle $a_\text{bf}$. The oracle is used only on small inputs, where exhaustive or clearly correct slow methods are feasible. We compare $\hat{a}^*$ and $a_\text{bf}$ on at least 1000 random small inputs. A candidate passes this gate only if the outputs match, the output format is deterministically judgeable, and $\hat{a}^*$ satisfies the target time complexity $\hat{T}^*$.

\paragraph{Gate 3: Similarity Filtering.}
We measure similarity between the source problem $q$ and the generated problem $q'$ using BM25 similarity, sentence-transformer cosine similarity, and $n$-gram overlap. These signals are combined into $\text{TextSim}(q,q')$. A candidate is rejected if
\begin{equation}
\text{TextSim}(q,q') \geq \tau_\text{text},
\end{equation}
where $\tau_\text{text}=0.55$ is calibrated using human annotation. We also require an algorithmic change, namely
\begin{equation}
\hat{\alpha} \neq \alpha
\quad\text{or}\quad
\hat{T}^* \neq T^*.
\end{equation}
This gate removes near-paraphrases and keeps problems whose source-to-target change is algorithmically meaningful.

\paragraph{Gate 4: Complexity Verification.}
Finally, the new reference solution $\hat{a}^*$ is checked by the deterministic complexity verifier described in Section~\ref{sec:verifier}. A candidate is rejected if the verifier marks the reference solution as \textsc{Uncertain} or reports a mismatch with the target time or space complexity. This gate ensures that every accepted problem has a certified algorithmic target.
\section{Example Benchmark Problems}
\label{app:examples}

\subsection*{Example 1: Range Query Shift (CS + SD)}

\noindent\colorbox{lightgray}{\parbox{0.96\linewidth}{
\textbf{Source.}
Given a static array of $n \leq 10^5$ integers, answer $Q \leq 10^5$ range-sum queries by outputting $\sum_{i=l}^{r} A[i]$. The reference solution uses prefix sums with $O(n+Q)$ time and $O(n)$ space.
}}

\noindent\colorbox{lightblue}{\parbox{0.96\linewidth}{
\textbf{Shifted variant.}
Given an array of $n \leq 2 \times 10^5$ integers and $Q \leq 2 \times 10^5$ operations, support either \texttt{1 l r x}, which adds $x$ to all elements in $A[l..r]$, or \texttt{2 l r}, which outputs $\max_{i=l}^{r} A[i]$. The new reference solution uses a lazy segment tree with $O((n+Q)\log n)$ time and $O(n)$ space.
}}

\noindent\colorbox{lightgray}{\parbox{0.96\linewidth}{
\textbf{Old-solution failure.}
The prefix-sum solution fails by both TLE and WA, since it cannot support range-add updates or range-maximum queries.
}}

\subsection*{Example 2: Greedy-Trap Interval Scheduling (GT)}

\noindent\colorbox{lightgray}{\parbox{0.96\linewidth}{
\textbf{Source.}
Given a set of intervals, select the maximum number of non-overlapping intervals. The reference solution uses the standard greedy rule based on earliest finishing time and runs in $O(n\log n)$ time.
}}

\noindent\colorbox{lightblue}{\parbox{0.96\linewidth}{
\textbf{Shifted variant.}
Each interval is assigned a type $t_i$. The goal is to select the maximum number of non-overlapping intervals while allowing at most $k$ consecutive type switches. The new reference solution uses dynamic programming over sorted intervals with a switch-count state and runs in $O(nk\log n)$ time.
}}

\noindent\colorbox{lightgray}{\parbox{0.96\linewidth}{
\textbf{Old-solution failure.}
The greedy solution fails by WA: it maximizes the number of intervals locally, but it does not account for the global switch-budget constraint.
}}

\subsection*{Example 3: Dynamic Connectivity Shift (SD)}

\noindent\colorbox{lightgray}{\parbox{0.96\linewidth}{
\textbf{Source.}
Given $n \leq 2000$ nodes and $m \leq 5000$ edges, answer connectivity queries. The reference solution uses DSU and runs in $O(n+m)$ time for the static graph setting.
}}

\noindent\colorbox{lightblue}{\parbox{0.96\linewidth}{
\textbf{Shifted variant.}
Given $n \leq 2 \times 10^5$ nodes and a sequence of $Q$ operations, support edge insertion, edge deletion, and connectivity queries. The new reference solution uses offline dynamic connectivity with DSU rollback and a segment tree over time, running in $O((n+Q)\log Q \cdot \alpha(n))$ time.
}}

\noindent\colorbox{lightgray}{\parbox{0.96\linewidth}{
\textbf{Old-solution failure.}
The vanilla DSU solution fails by TLE because it cannot handle deletions directly; rebuilding the DSU after deletions costs $O(Q(n+m))$ time.
}}

\section{Complexity Verifier: Implementation Details}
\label{app:verifier}

\paragraph{Static AST analysis.}
Python submissions are parsed with the \texttt{ast} module, while C++ submissions are parsed using LibClang Python bindings. The static analyzer tracks four types of evidence:
\begin{itemize}[noitemsep,leftmargin=*]
  \item nested loop depth over input-size parameters such as $N$, $Q$, and $M$, with loop bounds combined to estimate time complexity;
  \item recursion depth and branching factor, with memoized recursion handled separately;
  \item allocation sizes for arrays, vectors, and DP tables to estimate space complexity;
  \item data-structure signatures from class names, function-call patterns, and structural patterns, such as midpoint splitting in a recursive function as evidence for a segment tree.
\end{itemize}

\paragraph{Algorithm-tag detection rules.}
The verifier assigns algorithm tags using deterministic structural rules. Examples include:
\begin{itemize}[noitemsep,leftmargin=*]
  \item \texttt{lazy\_segment\_tree}: recursive range-query or range-update functions with lazy propagation before recursive descent;
  \item \texttt{dsu\_rollback}: union-find data structure with a history stack and explicit push/pop rollback operations;
  \item \texttt{dp\_with\_switch\_budget}: a 2D DP table where one dimension corresponds to a switch-count variable;
  \item \texttt{prefix\_sum\_only}: prefix-array allocation and range-sum formula without per-query dynamic updates.
\end{itemize}

\paragraph{Runtime scaling protocol.}
The verifier runs each solution on inputs with
$N \in \{10^3, 2\times10^3, 4\times10^3, 8\times10^3, 1.6\times10^4, 3.2\times10^4\}$.
For each input size, it performs five runs and uses the median runtime. It then fits a power-law model by log-log linear regression:
\begin{equation}
\log T = \alpha \log N + \gamma \log\log N + c.
\end{equation}
A solution is flagged as too slow if $\hat{\alpha} > \alpha_\text{target} + 0.15$, where the tolerance accounts for logarithmic factors and measurement noise.

\section{Prompt Templates}
\label{app:prompts}

We present the full prompt templates used for each evaluation strategy.
Placeholders in \texttt{\{curly braces\}} are filled per problem instance.
All prompts are delivered as multi-turn chat messages.

\subsection*{Strategy 1 — Direct (Zero-Shot)}

\begin{tcolorbox}[strategybox,title={Direct}]

\begin{tcolorbox}[systemprompt]
You are an expert competitive programmer.
Solve the algorithmic problem below by writing a complete, correct,
and efficient Python solution. Output only the code, no explanation.
\end{tcolorbox}

\begin{tcolorbox}[userprompt]
\textbf{Problem:} \{problem\_statement\}\\[2pt]
\textbf{Input format:} \{input\_format\}\\[2pt]
\textbf{Output format:} \{output\_format\}\\[2pt]
\textbf{Constraints:} \{constraints\}\\[2pt]
\textbf{Examples:}\\
\{examples\}\\[4pt]
Write a complete Python solution.
\end{tcolorbox}

\end{tcolorbox}

\subsection*{Strategy 2 — Chain-of-Thought (CoT)}

\begin{tcolorbox}[strategybox,title={Chain-of-Thought}]

\begin{tcolorbox}[systemprompt]
You are an expert competitive programmer who reasons carefully
before coding. Always reason step by step before writing code.
\end{tcolorbox}

\begin{tcolorbox}[userprompt]
\textbf{Problem:} \{problem\_statement\}\\[2pt]
\textbf{Input format:} \{input\_format\}\\[2pt]
\textbf{Output format:} \{output\_format\}\\[2pt]
\textbf{Constraints:} \{constraints\}\\[2pt]
\textbf{Examples:}\\
\{examples\}\\[4pt]
Reason step by step:\\
(1) Identify the key constraint that determines the required
    time complexity.\\
(2) Determine what time/space complexity class is needed.\\
(3) Select an algorithm or data structure that meets this class.\\
(4) Verify that the selected algorithm handles all edge cases.\\
(5) Implement the solution in Python.\\[2pt]
Output your reasoning followed by the complete Python code.
\end{tcolorbox}

\begin{tcolorbox}[assistprompt]
\textit{Step 1:} The constraint $N \leq 2 \times 10^5$ and $Q \leq
2\times 10^5$ updates/queries rule out any $O(NQ)$ approach$\ldots$\\
\textit{Step 2:} We need $O((N+Q)\log N)$.\\
\textit{Step 3:} A lazy segment tree supports range-add and
range-max in $O(\log N)$ per operation$\ldots$\\
\texttt{[Python code follows]}
\end{tcolorbox}

\end{tcolorbox}

\subsection*{Strategy 3 — Self-Refine}

Self-Refine \citep{madaan2023self} runs up to $R{=}3$ rounds.
Round 1 uses the Direct prompt; subsequent rounds append the
previous solution and a structured feedback message.

\begin{tcolorbox}[strategybox,title={Self-Refine — Feedback Round}]

\begin{tcolorbox}[systemprompt]
You are an expert competitive programmer.
You are given a problem, your previous solution, and feedback.
Revise the solution to fix all identified issues.
Output only the revised code.
\end{tcolorbox}

\begin{tcolorbox}[userprompt]
\textbf{Problem:} \{problem\_statement\}\\[2pt]
\textbf{Constraints:} \{constraints\}\\[4pt]
\textbf{Your previous solution:}\\
\texttt{\{previous\_solution\}}\\[4pt]
\textbf{Feedback:}\\
\{feedback\_message\}\\[2pt]
Please fix all issues and output a corrected Python solution.
\end{tcolorbox}

\begin{tcolorbox}[assistprompt,title={\sffamily\bfseries\footnotesize Feedback message (auto-generated)}]
Issue 1 (\textbf{Wrong Answer}): Your solution failed on hidden test
case 7. Expected output: \texttt{14}, your output: \texttt{12}.\\[2pt]
Issue 2 (\textbf{Complexity}): Static analysis detected a nested loop
over $N$ inside each query, giving $O(NQ)$. The target is
$O((N+Q)\log N)$. Replace the linear scan with a lazy segment tree.
\end{tcolorbox}

\end{tcolorbox}

\subsection*{Strategy 4 — RAG-Source Retrieval}

RAG-source retrieves the top-1 most similar source problem
(by BM25 + embedding re-rank) and prepends it to the Direct prompt.
The retrieval corpus consists of all seed problems in \sys
with their original reference solutions.

\begin{tcolorbox}[strategybox,title={RAG-Source}]

\begin{tcolorbox}[systemprompt]
You are an expert competitive programmer.
You will be shown a related reference problem and its solution,
followed by a new problem with modified constraints.
\textbf{Important:} the new problem has different requirements.
Do NOT copy the reference solution; adapt it if necessary.
\end{tcolorbox}

\begin{tcolorbox}[userprompt]
\textbf{Reference problem (similar, but different):}\\
\{retrieved\_source\_statement\}\\[2pt]
\textbf{Reference solution:}\\
\texttt{\{retrieved\_source\_solution\}}\\[4pt]
\rule{\linewidth}{0.4pt}\\[4pt]
\textbf{New problem (solve this one):}\\
\{problem\_statement\}\\[2pt]
\textbf{Input format:} \{input\_format\}\\[2pt]
\textbf{Output format:} \{output\_format\}\\[2pt]
\textbf{Constraints:} \{constraints\}\\[2pt]
\textbf{Examples:}\\
\{examples\}\\[4pt]
The constraints have changed. Determine whether the reference
solution still applies or a new algorithm is required.
Write a complete Python solution.
\end{tcolorbox}

\end{tcolorbox}

\subsection*{Strategy 5 — Skill-Guided}

Skill-guided augments the CoT prompt with explicit complexity
targets and forbidden patterns derived from the problem's
complexity metadata (Section~\ref{sec:verifier}).

\begin{tcolorbox}[strategybox,title={Skill-Guided}]

\begin{tcolorbox}[systemprompt]
You are an expert competitive programmer.
You will be given a problem together with the required
time/space complexity and the expected algorithmic approach.
Use this information to guide your solution.
\end{tcolorbox}

\begin{tcolorbox}[userprompt]
\textbf{Problem:} \{problem\_statement\}\\[2pt]
\textbf{Input format:} \{input\_format\}\\[2pt]
\textbf{Output format:} \{output\_format\}\\[2pt]
\textbf{Constraints:} \{constraints\}\\[2pt]
\textbf{Examples:}\\
\{examples\}\\[4pt]
\rule{\linewidth}{0.4pt}\\[4pt]
\textbf{Complexity guidance:}\\
$\bullet$ \textbf{Required time complexity:} \{target\_time\_complexity\}\\
$\bullet$ \textbf{Required space complexity:} \{target\_space\_complexity\}\\
$\bullet$ \textbf{Expected algorithm class:} \{expected\_algorithm\_tags\}\\
$\bullet$ \textbf{Avoid:} \{forbidden\_algorithm\_tags\} — these are
  too slow or incorrect under the given constraints.\\[4pt]
Implement an efficient Python solution that satisfies the above
complexity requirements. Think step by step.
\end{tcolorbox}

\end{tcolorbox}

\subsection*{Strategy 6 — Reflexion}

Reflexion \citep{shinn2023reflexion} maintains a persistent verbal
memory of past failures. At each round the model first writes a
short reflection on why its previous attempt failed, stores it,
and then generates a new solution conditioned on the memory.

\begin{tcolorbox}[strategybox,title={Reflexion}]

\begin{tcolorbox}[systemprompt]
You are an expert competitive programmer with a memory of past
attempts. Learn from your previous failures and improve.
\end{tcolorbox}

\begin{tcolorbox}[userprompt]
\textbf{Problem:} \{problem\_statement\}\\[2pt]
\textbf{Constraints:} \{constraints\}\\[2pt]
\textbf{Examples:}\\
\{examples\}\\[4pt]
\textbf{Memory of past attempts:}\\
\{reflexion\_memory\}\\[4pt]
Based on your memory, write a new Python solution that avoids
the mistakes you identified. Be explicit about what you are
changing and why.
\end{tcolorbox}

\begin{tcolorbox}[assistprompt,title={\sffamily\bfseries\footnotesize Example reflexion memory entry}]
\textit{Attempt 1 reflection:} I used a prefix sum array to answer
range queries, but I forgot that this problem also has range-add
updates. Prefix sums become stale after any update. I need a
data structure that handles both updates and queries efficiently
--- a lazy segment tree or a Fenwick tree with range operations.
Next time: check for the presence of update operations before
choosing a static data structure.
\end{tcolorbox}

\end{tcolorbox}
\end{document}